\documentclass[aps,twocolumn,prd,
		showpacs,preprintnumbers,amsmath,amssymb,nofootinbib]{revtex4-1}

\pdfoutput=1

\usepackage{hyphenat}
\usepackage[caption=false]{subfig}
\usepackage{graphicx}
\usepackage{mathtools}				% for \DeclarePairedDelimiter (s.u.)
\usepackage{relsize}
\usepackage{xcolor}
\usepackage{dsfont}
\usepackage{units}					% for \nicefrac
\usepackage{csquotes}					% for \enquote
\usepackage{siunitx}
\usepackage{xspace}
\usepackage{multirow}

\usepackage{hyperref}

\newcommand{\eg}{e.g.~}
\newcommand{\cf}{cf.~}
\newcommand{\ie}{i.e.~}
\newcommand{\CL}{C.L.~}

\newcommand{\symhspace}[2]{\hspace{#1}#2\hspace{#1}}	% symmetrisches Einfuegen von whitespace

\newcommand{\I}{\mathrm{i}}					% imaginäre Einheit
\newcommand{\fineq}[1]{\;{#1}}				% Satzzeichen am Ende von abgesetzten Formeln
\newcommand{\sepeq}[1]{\;{#1} \quad}	% Satzzeichen zwischen abgesetzten Formeln
\newcommand{\dd}{\mbox{d}}				% Differential-d
\DeclarePairedDelimiter\absVal{\vert}{\vert}

\newcommand{\Bmumu}{\mathcal{B}_{\mu\mu}}

\newcommand{\figPath}{\string.}

\definecolor{darkblue}{rgb}{0.0,0.0,0.4}
\definecolor{darkgreen}{rgb}{0.0,0.4,0.0}
\hypersetup{
    colorlinks,
    linkcolor=darkblue,
    citecolor=darkgreen,
    urlcolor=darkblue
}

\begin{document}

%%%%%%%%%%%%%%%%%%%%%%%%%%%%%%%%
%%%%%%%%  T I T L E   &   A U T H O R S   %%%%%%%%%
%%%%%%%%%%%%%%%%%%%%%%%%%%%%%%%%

\title{Prospects for three-body Higgs boson decays into extra light scalars}

\author{Alexander J. Helmboldt}
\thanks{Corresponding author}
\email{alexander.helmboldt@mpi-hd.mpg.de}

\author{Manfred Lindner}

\affiliation{\vspace{0.5em}Max-Planck-Institut f\"ur Kernphysik, Saupfercheckweg 1, 69117
  Heidelberg, Germany}


%%%%%%%%%%%%%%%%%%%%%%%%%%%%%%%%
%%%%%%%%  A B S T R A C T  %%%%%%%%%%%%%%%
%%%%%%%%%%%%%%%%%%%%%%%%%%%%%%%%

\begin{abstract}
\noindent
%%%%% Introduction
Within models containing a very light scalar particle coupled to the \SI{125}{GeV} Higgs boson, we present the first detailed study of Higgs decays into \textit{three} of these light scalars.
%%%%% Model-independent scenarios with large \Gamma_3
We determine model-independent conditions which the scalar sector after electroweak symmetry breaking has to satisfy in order for the three-body channel to become relevant.
%%%%% Model-specific analysis (conditions for large \Gamma_3)
Using a specific model \textendash\ the real scalar singlet-extension of the Standard Model (SM) \textendash\ we then identify scenarios, where the rates of scalar three-body Higgs decays are comparable to or even exceed those of the well-studied two-body channel.
All those scenarios are shown to be compatible with current experimental and theoretical constraints.
%%%%% Model-specific analysis (collider signatures)
We finally argue that scalar three-body Higgs decays lead to exciting new collider signatures with \textit{six} SM fermions in the final state. Calculating the corresponding event rates, we find that \eg six-muon or six-tau final states may be in reach of dedicated searches at the LHC or ILC experiments.
\end{abstract}

\maketitle

%%%%%%%%%%%%%%%%%%%%%%%%%%%%%%%%
%%%%%%%%  S E C T I O N :  Introduction %%%%%%%%%
%%%%%%%%%%%%%%%%%%%%%%%%%%%%%%%%

\section{Introduction}
\label{sec:intro}
\noindent
%================================================
With the discovery of the Higgs boson at the Large Hadron Collider (LHC) \cite{Aad2012b,Chatrchyan2012a}, the last missing piece of the Standard Model (SM) of particle physics was confirmed.
Nevertheless, it is not yet clear whether the found scalar is precisely the one predicted by the SM.
Accordingly, one of the most important goals of current and future particle physics experiments is to accurately measure the Higgs boson's properties in order to clarify whether the SM description of electroweak symmetry breaking is complete.
However, there are multiple reasons to think that this is not the case and the scalar sector is not minimal.

For instance, particle physics models which try to explain the origin of the gauge-hierarchy problem \textendash\ \ie the two-fold question of why the Higgs mass can be small and radiatively stable in the presence of some high-energy embedding of the SM \textendash\ often introduce new scalar degrees of freedom.
As a prime example, we mention supersymmetric extensions of the SM which necessarily enlarge the Higgs sector by at least a second complex scalar doublet.
But also other approaches as little Higgs models or theories based on scale invariance and Coleman-Weinberg symmetry breaking inevitably exhibit an augmented scalar sector.

There are also other shortcomings of the minimal SM which motivate postulating additional scalar particles. For example, establishing a link between the origin of the baryon asymmetry of the Universe and electroweak physics in theories of electroweak baryogenesis requires the electroweak phase transition to be strongly first-order. Whereas the SM fails to provide such a transition, it can be realized by appropriately extending the model's scalar particle content.
Another motivation for a nonminimal Higgs sector are particle physics models of cosmological inflation. These rely on the existence of a scalar field and its associated excitations, the inflaton. If the inflaton is different from the SM Higgs, these theories necessarily predict an additional scalar particle.
Last but not least, there exists a plethora of models of particle dark matter in which new scalar degrees of freedom constitute the dark sector.

In the present work, we will mainly be interested in additional scalar particles that are much lighter than the SM-like Higgs boson found at the LHC. Such light scalars are predicted, for instance, in models of Coleman-Weinberg symmetry breaking, where they naturally arise as the pseudo-Goldstone bosons of the anomalously broken scale invariance, see \eg \cite{Gildener1976,Foot2007}.
As a second example, let us mention the next-to-minimal supersymmetric SM, in which a light pseudoscalar particle appears naturally (for recent reviews, see \eg \cite{Maniatis2010,Ellwanger2010a}).

Irrespective of the particular motivation, models with an extended scalar sector have several things in common. Most importantly, as soon as one extends the SM by another scalar field $S$, the most general, renormalizable Lagrangian inevitably contains a scalar portal to the complex Higgs doublet $\Phi$
\begin{align*}
	\mathcal{L} \supseteq - \lambda_p (\Phi^\dagger \Phi) (S^\dagger S) \fineq{.}
\end{align*}
If the new degrees of freedom are sufficiently light, the above portal necessarily induces the decay of the physical \SI{125}{GeV} Higgs boson $H$ to two new scalars,
\begin{align}
	H \to SS \fineq{.}
	\label{eq:intro:twoBodyDecay}
\end{align}
Depending on the exact properties of $S$, this decay may manifest itself in different ways.
On the one hand, if $S$ is sufficiently stable and thus decays only outside of the detector (if at all), scalar Higgs decays cannot be observed directly. Still, the aforementioned process will contribute to the invisible Higgs width and hence modify the signal strengths of Higgs decays into SM particles, which are currently measured at the LHC.
The same logic applies if $S$ decays rather quickly, but predominantly into further hidden-sector particles that are undetectable in experiments.
On the other hand, in cases where interactions of the new scalar and SM particles are sufficiently large, the process in \eqref{eq:intro:twoBodyDecay} can give interesting collider signatures with SM particles in the final state, but characteristic features such as displaced vertices or multiple lepton jets, see \eg \cite{Curtin2014}.
The above discussion demonstrates that, apart from being well motivated from a theoretical point of view, light scalars are experimentally extremely promising.
Accordingly, the phenomenological implications of the two-body Higgs decay in \eqref{eq:intro:twoBodyDecay} have been extensively studied in different contexts, see \eg \cite{Gunion2000,Curtin2014} and references therein.

At the same time, to our best knowledge, the corresponding three-body decay channel of the LHC Higgs,
\begin{align}
	H\to SSS \fineq{,}
	\label{eq:intro:threeBodyDecay}
\end{align}
has never been discussed in the literature before.
However, this process is quite common as the majority of models with enlarged scalar sector predict it if kinematically allowed.
These models include, but are not restricted to theories where the SM physical Higgs degree of freedom mixes with some other $CP$-even scalar which can, of course, be part of a larger electroweak multiplet.

With the present paper, we attempt to fill the apparent gap in the literature pointed out before.
To be more precise, we will show that there exist scenarios, consistent with current experimental and theoretical constraints, in which the aforementioned three-body rate can become comparable to or even exceed that of two-body scalar Higgs decays.
In particular, we will formulate the physical requirements for this to happen as model-independent as possible.
From a more phenomenological perspective, we will demonstrate that three-body scalar Higgs decays may give rise to unique and very clean signatures with non-negligible rates at the LHC and future electron-positron colliders.
Furthermore, we will argue that the search for such decays can provide a method to distinguish different beyond-the-SM theories or to constrain a given model's scalar sector.

Reflecting the above outline, the article is organized as follows. In Section \ref{sec:decays}, we discuss scalar two- and three-body decays of the LHC Higgs for a generic low-energy scalar sector. We then specialize to a simple model in which the aforementioned low-energy scalar sector is realized, namely the SM extended by a real scalar singlet (Section \ref{sec:SMS}). Furthermore, we study in that section relevant constraints on the model's parameter space.
In Section \ref{sec:pheno} we then see how the above constraints set limits on both two- and three-body decay rates as well as on those rates' ratio. We analyze under which circumstances three-body Higgs decays can become relevant and their possible collider signatures.
We finally summarize our findings in Section \ref{sec:concl}.

%%%%%%%%%%%%%%%%%%%%%%%%%%%%%%%
%%%%%%%%  S E C T I O N :  Scalar Higgs decays %%%%
%%%%%%%%%%%%%%%%%%%%%%%%%%%%%%%

\section{Scalar Higgs decays}
\label{sec:decays}
\noindent
%================================================
In the present section, we will start by considering a \textit{generic} low-energy effective theory which is assumed to describe physics after spontaneous electroweak symmetry breaking. This theory's spectrum is supposed to contain at least two electrically neutral and colorless, physical scalar particles $H$ and $h$, whose mutual and self-interactions are governed by the following potential
\begin{align}
	\begin{split}
		V(H,h) ={}& \tfrac{m_H^2}{2} H^2 + \tfrac{m_h^2}{2} h^2 + \lambda_{4H} H^4 + \lambda_{4h} h^4 \\
		& + \kappa_{3H} H^3  + \kappa_{3h} h^3 + \kappa_{H2h} H h^2 + \kappa_{2Hh} H^2 h \\
		& + \lambda_{2H2h} H^2 h^2 + \lambda_{3Hh} H^3 h + \lambda_{H3h} H h^3 \fineq{,}
	\end{split}
	\label{eq:decays:potential}
\end{align}
with dimensionless parameters $\lambda_i$ and trilinear couplings $\kappa_i$ of mass dimension one.
In the rest of this work, we will be interested in the situation, where the heavier scalar $H$ is identified with the LHC Higgs boson, whereas $h$ stems from some hidden sector. The discussion in this section, however, is independent of this association.
Note that depending on the underlying model's symmetries and particle content, one or several of the above effective couplings might be exactly zero.

Now, let the scalar masses satisfy $m_H \geq 3 m_h$, such that the decays $H\to2h$ and $H\to3h$ are both kinematically allowed.
Defining the ratio of scalar masses $x := m_h/m_H$, the two-body decay rate at tree-level is given by
\begin{align}
	\Gamma_2 \equiv \Gamma(H\to2h) = \frac{\kappa_{H2h}^2}{8\pi m_H} \cdot \ell_2(x) \fineq{,}
	\label{eq:decays:2BodyTree}
\end{align}
where the kinematic threshold function is well-known, $\ell_2(x) = \sqrt{1-4x^2}$.

\begin{figure}[t]
	\centering
	\subfloat[]{\includegraphics[scale=0.36]{\figPath/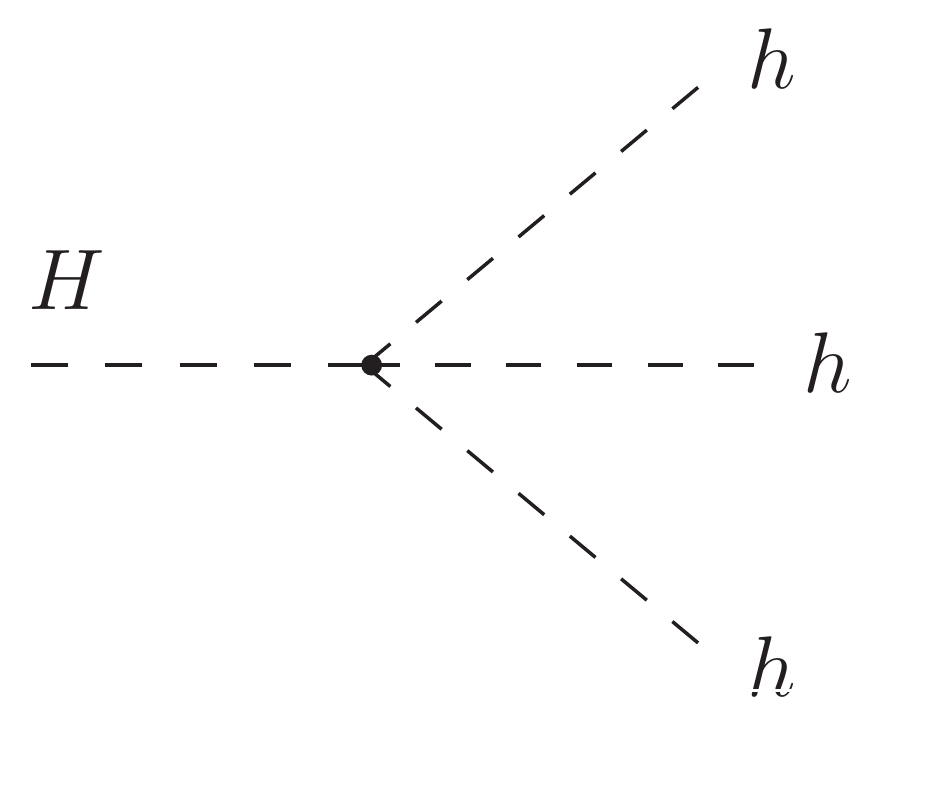}}%
	\hspace{2em}%
	\subfloat[]{\includegraphics[scale=0.36]{\figPath/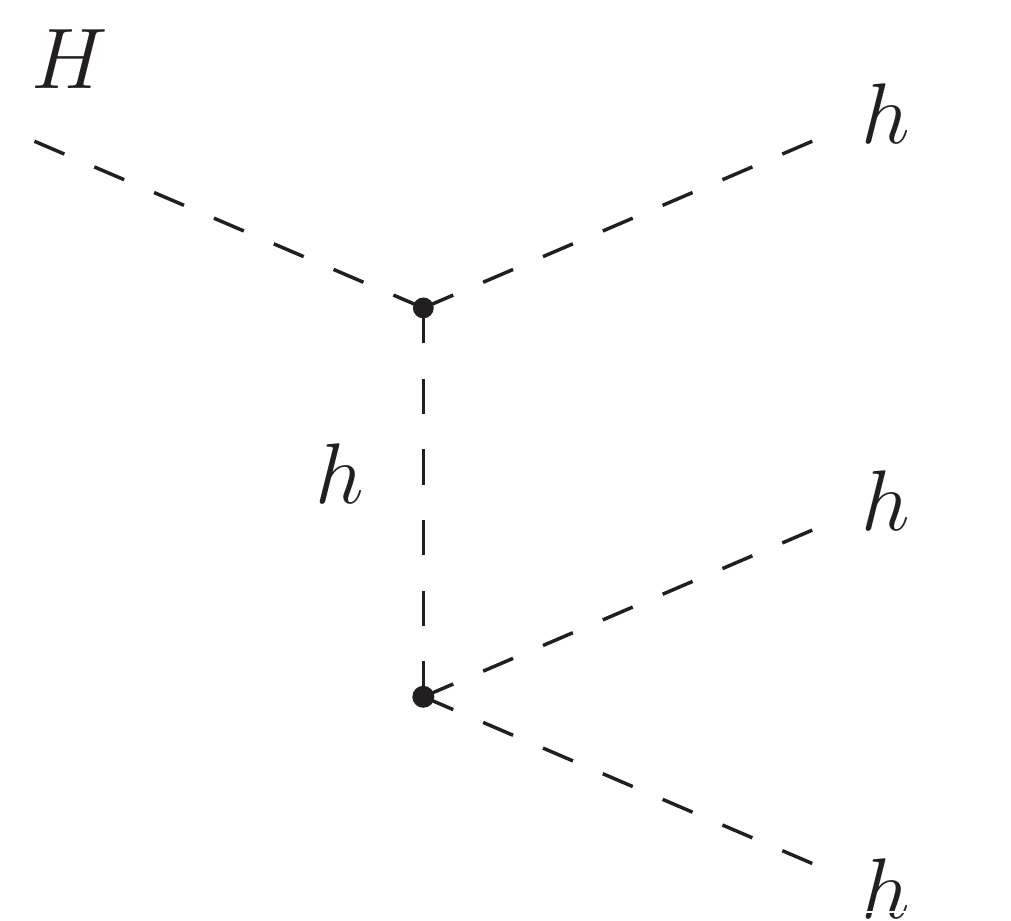}}%
	\caption{Feynman graphs contributing to the decay $H\to 3h$ at tree-level: (a) contact interaction and (b) $h$-exchange. For (b) there exist two further final-state permutations which are not shown.}
	\label{fig:decays:3BodyTree}
\end{figure}

The Lagrangian \eqref{eq:decays:potential} also entails the scalar three-body decay of $H$, which is mediated by two types of processes at tree-level: by a contact interaction proportional to $\lambda_{H3h}$ and by the exchange of a virtual light scalar. The corresponding Feynman graphs are shown in Figure \ref{fig:decays:3BodyTree}.
An explicit calculation yields
\begin{align}
	\begin{split}
		\Gamma_3 = \frac{9 m_H}{64 \pi^3} \biggl[ &
		\frac{\lambda_{H3h}^2}{12} \ell^{(0)}_3(x) 
		+ \lambda_{H3h}\frac{\kappa_{H2h}\kappa_{3h}}{m_H^2} \ell^{(1)}_3(x) \\
		& + \frac{\kappa_{H2h}^2\kappa_{3h}^2}{m_H^4}
		\left( \ell^{(2)}_3(x) + 2 \ell_3^{(1,1)}(x) \right) \biggr] \fineq{.}
	\end{split}
	\label{eq:decays:3BodyTree}
\end{align}
The first and third term in the above expression stem from the squared contact and $h$-exchange interaction graphs, respectively. The second term describes interference of contact and $h$-mediated contributions whereas the fourth term contains interference between $h$-exchange diagrams of different permutations in the final state. The various three-body threshold functions $\ell_3$ are discussed in Appendix \ref{app:phaseSpace}.

As a measure to quantify the three-body channel's importance relative to the two-body channel, it is useful to define the ratio of partial widths 
\begin{align}
	r := \frac{\Gamma_3}{\Gamma_2} \equiv \frac{\mathcal{B}(H \to 3h)}{\mathcal{B}(H \to 2h)} \fineq{.}
	\label{eq:decays:ratio}
\end{align}
In the present work, we will be particularly interested in scenarios where $r$ can become of order one or even larger.
Naively, however, one would expect $r \ll 1$ for several reasons. On the one hand, the three-body final state will have a smaller phase space. The associated suppression decreases, however, as $m_h/m_H$ becomes tiny.
On the other hand, diagrams containing additional internal propagators will be suppressed by an extra coupling and by the virtual particle's mass. Because of the latter, we neglect graphs with a virtual $H$ in equation \eqref{eq:decays:3BodyTree} and Figure \ref{fig:decays:3BodyTree}. For the $h$-mediated diagrams, the above effect is minimized provided $h$ is sufficiently light and $\kappa_{3h}$ is large enough. Obviously, this kind of suppression is absent for the contact interaction.

In Sections \ref{sec:SMS} and \ref{sec:pheno}, we will study under which circumstances $r\simeq 1$ is possible in a specific model.
With this in mind, it is helpful to revisit equations \eqref{eq:decays:2BodyTree} and \eqref{eq:decays:3BodyTree} and identify the significance of the individual low-energy effective couplings.
That way, the requirements for $r$ to be sizable can be formulated as model-independent as possible. Then, in order to find the parameters crucial for the size of $r$ in a given theory, one only needs to determine how the relevant effective couplings depend on that model's fundamental parameters.

From equations \eqref{eq:decays:2BodyTree} and \eqref{eq:decays:3BodyTree} we now see that the scalar $H$ decay widths at tree-level are governed by only three couplings, namely $\kappa_{H2h}$, $\kappa_{3h}$ and $\lambda_{H3h}$.
More precisely, the trilinear portal $\kappa_{H2h}$ will determine the overall size of the scalar two-body as well as that of the $h$-mediated three-body rates.
The ratio of these two rates is fixed by the hidden sector's trilinear self-interaction $\kappa_{3h}$.
Lastly, the contribution of contact interactions to the three-body width is set by $\lambda_{H3h}$.
Already this brief discussion demonstrates that $\Gamma_3$ can become comparable to $\Gamma_2$ if the light scalar's cubic self-interactions are sufficiently strong.
Small $\kappa_{H2h}$ together with a large enough $\lambda_{H3h}$ constitutes a different way to realize sizable ratios $r$.
We will learn more about the relative importance of the two aforementioned effects when discussing a specific model in Section \ref{sec:pheno}.

Finally, let us stress an important conceptual difference between scalar two- and three-body decays: On the one hand, measuring $\Gamma_2$ amounts to exclusively test the coupling between the two scalar sectors.
On the other hand, studying $\Gamma_3$ gives the possibility of quantifying the hidden-sector self-interactions which might otherwise be inaccessible at colliders.%
\footnote{Strictly speaking, the above reasoning is only true at tree-level since already at one-loop the process $H\to hh$ obtains corrections involving both $\kappa_{3h}$ and $\lambda_{4h}$. However, those corrections will typically be very small and thus hard to observe.}

%%%%%%%%%%%%%%%%%%%%%%%%%%%%%%%
%%%%%%%%  S E C T I O N :  Singlet-extended SM %%%%
%%%%%%%%%%%%%%%%%%%%%%%%%%%%%%%

\section{Scalar singlet-extension of the Standard Model}
\label{sec:SMS}
\noindent
%================================================
After having discussed generic properties of multibody scalar Higgs decays in the last section, let us now study a specific particle physics model which leads to a scalar potential of the form \eqref{eq:decays:potential} at low energies. As a working example, we consider the minimal extension of the Standard Model (SM) in which the Higgs sector is supplemented by one real scalar gauge singlet $S$ (see \eg  \cite{Silveira1985,Krasnikov1992,Schabinger2005,Patt2006,OConnell2007}).
Without imposing any additional symmetry, the model's most general renormalizable potential before electroweak symmetry breaking can be parametrized as
\begin{align}
	\begin{split}
		V(\Phi,S) ={}& \frac{\mu^2}{2} \Phi^\dagger\Phi + \lambda (\Phi^\dagger\Phi)^2 \\
		&+ \frac{\delta_1}{2} (\Phi^\dagger\Phi) S + \frac{\delta_2}{2} (\Phi^\dagger\Phi) S^2 \\
		&+ \kappa_1 S + \frac{\kappa_2}{2} S^2 + \frac{\kappa_3}{3} S^3 + \frac{\kappa_4}{4} S^4 \fineq{,}
	\end{split}
	\label{eq:SMS:potential}
\end{align}
where $\Phi$ is the usual complex Higgs doublet. In unitary gauge, its neutral component after electroweak symmetry breaking can be written as
\begin{align}
	\Phi^0 = \frac{v+\phi}{\sqrt{2}} \quad \sepeq{\text{with}} v=\SI{246}{GeV} \fineq{.}
	\label{eq:SMS:vevStructure}
\end{align}
Due to the term linear in $S$ the singlet's vacuum expectation value can be chosen to vanish without loss of generality. The corresponding solutions of the model's tadpole equations then read $\kappa_1 = -\delta_1 v^2/4$ and $\mu^2 = -\lambda v^2/2$.

Importantly, since both $\phi$ and $S$ are neutral and colorless, the $\delta_1$-term will in general lead to mixing, the strength of which can be parametrized by one real angle $\theta$. The scalar mass eigenstates can then be written as
\begin{align}
	\begin{split}
		H & = \cos\theta \cdot \phi + \sin\theta \cdot S \fineq{,} \\
		h & = \cos\theta \cdot S - \sin\theta \cdot \phi \fineq{.}
	\end{split}
	\label{eq:SMS:mass_eigenstates}
\end{align}
In the rest of this paper, we will exclusively be interested in the situation where $h$ is much lighter than the Higgs boson $H$ found at the LHC, \ie $m_h \ll m_H = \SI{125}{GeV}$ \cite{Aad2015}.
As we will argue below, various experiments then require the mixing angle to be small.
Nevertheless, equations \eqref{eq:SMS:mass_eigenstates} demonstrate that the light, singletlike mass eigenstate $h$ will have the \textit{same} couplings as the SM Higgs, but suppressed by an additional factor of $\sin\theta$. Likewise, the couplings of the heavy Higgs $H$ will be slightly modified with respect to the pure SM by $\cos\theta$.

Diagonalizing the model's scalar mass matrix can be done analytically resulting in the known formulas for scalar masses and mixing angle in terms of Lagrangian couplings.
For the sake of clearness, it is helpful to invert these relations in order to trade some of the Lagrangian parameters in \eqref{eq:SMS:potential} for physical particle masses and mixing, namely
\begin{align}
	\begin{split}
		\lambda & = \frac{2}{v^2} \left(  m_H^2 \cos^2\theta + m_h^2 \sin^2\theta \right) \fineq{,} \\
		\kappa_2 & = -\frac{\delta_2 v^2}{2} +m_h^2 \cos^2\theta + m_H^2 \sin^2\theta \fineq{,} \\
		\delta_1 & = \frac{m_H^2 - m_h^2}{v} \sin 2\theta \fineq{.}
	\end{split}
	\label{eq:SMS:Lag2Phys}
\end{align}
Summarizing, the model's scalar sector is now characterized by three dimensionless parameters ($\sin\theta$, $\delta_2$, $\kappa_4$) and two couplings ($m_h$ and $\kappa_3$) with mass dimension one, supplemented by the known quantities $m_H=\SI{125}{GeV}$ and $v=\SI{246}{GeV}$.

Obviously, the scalar sector of the singlet-extended SM after electroweak symmetry breaking will be of the form studied in Section \ref{sec:decays}.
The exact relations between the generic couplings in \eqref{eq:decays:potential} and the aforementioned model parameters are obtained by plugging in equations \eqref{eq:SMS:vevStructure} to \eqref{eq:SMS:Lag2Phys} into the potential \eqref{eq:SMS:potential}.
In the next paragraph, we will argue that experimental bounds only allow for relatively small portal couplings $\delta_2$ as well as tiny scalar mixing angles. To simplify the calculations in the remainder of this paper, it is therefore helpful to expand the relevant effective scalar couplings to first order in $\theta$:
\begin{align}
	\begin{split}
		\kappa_{H2h} & \simeq \tfrac{1}{2}\delta_2 v + \kappa_3 \theta \sepeq{,} \quad \hspace{0.23em}
		\kappa_{3h} \simeq \tfrac{1}{3} \kappa_3 - \tfrac{1}{2} \delta_2 v \theta \fineq{,} \\
		\lambda_{H3h} & \simeq (\kappa_4 - \tfrac{1}{2}\delta_2) \theta \sepeq{,} \quad
		\lambda_{4h} \simeq \kappa_4/4 \fineq{.}
	\end{split}
	\label{eq:SMS:smallThetaExp}
\end{align}
Equations \eqref{eq:SMS:smallThetaExp} together with the results from Section \ref{sec:decays} now also allow us to identify the roles of the individual fundamental parameters. For simplicity, we thereby restrict ourselves to the decoupling limit $\theta\to 0$. The generalization to moderate values of the mixing angle is, however, straightforward.
In the decoupling limit, the quartic portal coupling $\delta_2$ fixes the value of $\kappa_{H2h}$ and hence the overall size of the scalar two-body as well as that of the $h$-mediated three-body Higgs decay rates.
The ratio of these two rates is governed by the trilinear scalar self-interaction, \ie by $\kappa_3$. Importantly, $\kappa_{3h}$ is \textit{not} suppressed by the mixing angle and therefore remains finite as $\theta$ tends to zero.
Since already consistency with experiment requires $\delta_2$ to be small, the value of $\lambda_{H3h}$ and thus the contribution of contact interactions to the three-body decay is mainly determined by $\kappa_4$ and $\theta$.
A lesson to be learned here is that for three-body decay rates to be sizable, the singlet sector has to exhibit relatively strong self-interactions, \ie $\kappa_3$ and $\kappa_4$ need to be large. Alternatively, not observing any three-body Higgs decays may be used to constrain said self-interactions.

%%%%%%%%  S U B S E C T I O N :  Constraints on the parameter space %%%%%%%%

\subsection{Constraints on the parameter space}
\label{sec:SMS:constr}
\noindent
%================================================
In this section, we will briefly address various experimental and theoretical constraints that narrow down the viable parameter space of the singlet-extended SM. We thereby concentrate on the couplings, which were identified to govern the scalar Higgs decays in Section \ref{sec:decays} and below equation \eqref{eq:SMS:smallThetaExp}.

For a first analysis, we will \textit{a priori} restrict ourselves to only a subset of possible light Higgs masses, which is chosen as to allow for promising signatures of scalar three-body Higgs decays at current or future colliders:
On the one hand, we will focus on masses above $(m_K-m_\pi) \approx \SI{360}{MeV}$ since for lighter scalars the mixing angle is already very tightly constrained by kaon decays \cite{Clarke2014} and the singlet sector virtually decouples.%
\footnote{Still, we will briefly discuss this decoupling limit in Section \ref{sec:pheno}.}
On the other hand, we will assume $m_h$ to lie below roughly $\SI{10}{GeV}$ because for larger masses phase space suppression of the three-body channel starts to become relevant.

%%%%%%%%  S U B S E C T I O N :  Experimental constraints %%%%%%%%

\subsubsection{Experimental constraints}
\noindent
%================================================
For light scalar masses above \SI{360}{MeV} and below the $B$-meson threshold at roughly \SI{5}{GeV}, the severest bound on the scalar mixing angle $\theta$ arises from the measurement of inclusive $B$-meson decays \cite{Clarke2014}. The relevant branching fractions have to satisfy \cite{Olive2014}
\begin{align*}
	\mathcal{B}(B \to X_s + h) \mathcal{B}(h \to \mu^+ \mu^-) & < \mathcal{B}(B \to X_s \ell^+ \ell^-)\\
	& = \left( 3.66^{+0.76}_{-0.77} \right) \cdot 10^{-6} \fineq{.}
\end{align*}
The process $B\to X_s +h$ is predominantly mediated by a penguin diagram and the corresponding branching ratio can be calculated in effective field theory \cite{Chivukula1988}. Using this, one obtains the following limit
\begin{align}
	\sin^2\theta \cdot \mathcal{B}(h\to\mu^+\mu^-) \lesssim \num{0.51e-6} \fineq{.}
	\label{eq:SMS:upperLimitTheta}
\end{align}
A constraint similar to that in equation \eqref{eq:SMS:upperLimitTheta} is obtained from the exclusive decay channel $B \to K \mu^+\mu^-$ \cite{Clarke2014}.
In any case, calculating the branching fraction of $h\to\mu^+\mu^-$ turns out to be rather complicated for a very light $h$, since nonperturbative QCD effects have to be taken into account properly in computing the light Higgs' total decay width.
We will discuss this issue in more detail in Section \ref{sec:pheno:hdecay}.
For now, it suffices to say that $\mathcal{B}(h\to\mu^+\mu^-)$ ranges between \SI{1}{\%} and \SI{10}{\%} in the interesting mass region.
Consequently, $\sin\theta$ cannot be larger than $10^{-2}$.

For light scalar masses above the $B$-meson threshold, the decay $B\to h+X_s$ is kinematically forbidden. Bounds on $\theta$ now come from LEP searches for the Bjorken process $e^+ e^- \to Zh$ \cite{Buskulic1993,Acciarri1996,Abbiendi2003} and $\Upsilon$ decays \cite{Wilczek1977d}. The former give the strongest constraints, namely $\sin\theta \lesssim \num{0.1}$ for $m_h$ below $\SI{10}{GeV}$ \cite{Clarke2014}.
The given limit thereby comes from a dedicated L3 analysis which assumed a hadronic decay of the produced light Higgs \cite{Acciarri1996}.
Bounds from a decay-mode- and thus model-independent analysis by OPAL are significantly weaker \cite{Abbiendi2003,Clarke2014}. However, since the light scalar in the singlet-extended SM behaves like the SM Higgs except for its rescaled couplings to SM particles, the stricter L3 numbers apply in our case.

A further important constraint on the model's parameter space comes from measuring Higgs signal strengths at the LHC \cite{Aad2016z}:
Observing that SM calculations describe the experimental findings very well sets an upper bound of $\mathcal{B}^\text{max}_\text{non}=\SI{34}{\%}$ at \SI{95}{\%} \CL on the Higgs branching fraction for decays into nonstandard final states \cite{Aad2016z}.
The stated number is obtained based on only a few assumptions,%
\footnote{Within the $\kappa$-framework, it is assumed that $\absVal{\kappa_\mathsmaller{W}}, \absVal{\kappa_\mathsmaller{Z}}\leq 1$, \mbox{$\kappa_\mathsmaller{W} \cdot \kappa_\mathsmaller{Z}>0$} and that coupling modifiers do not change when going from \SI{7}{TeV} to \SI{8}{TeV}. For more details, see \cite{Aad2016z}.}
with all of which the singlet-extended SM is compatible.
The associated maximally allowed nonstandard Higgs width is found to be
\begin{align}
	\Gamma^\text{max}_\text{non}
	= \cos^2\theta \cdot \Gamma_H^\text{SM} \frac{\mathcal{B}^\text{max}_\text{non}}{1-\mathcal{B}^\text{max}_\text{non}}
	\simeq \SI{2.1}{MeV} \fineq{.}
	\label{eq:SMS:GammaNonMax}
\end{align}
In order to obtain the numerical result, we used the SM Higgs total width $\Gamma_H^\text{SM}=\SI{4.088}{MeV}$ \cite{DeFlorianSabaris:2215893}, as well as the fact that scalar mixing must be small.
The nonstandard Higgs decays are now precisely those to multiple light scalars. If only the two-body rate is sizable, we get for $m_h\ll m_H$
\begin{align}
	\Gamma(H\to2h) \stackrel{!}{\leq} \Gamma^\text{max}_\text{non} 
	\sepeq{\quad\Rightarrow} \delta_2 \lesssim \num{0.021} \fineq{,}
	\label{eq:SMS:invisibleHiggs_delta2}
\end{align}
where we again employed the small-$\theta$ limit, in which $\kappa_{H2h} \simeq v\delta_2/2$.
Obviously, the above bound on $\delta_2$ will become even stronger in regions of parameter space where the three-body decay rate cannot be neglected.

%%%%%%%%  S U B S E C T I O N :  Theoretical constraints %%%%%%%%

\subsubsection{Theoretical constraints}
\label{sec:SMS:theoretical}
\noindent
%================================================
Complementary to what we did in the last paragraph, we will now consider constraints on the scalar sector of the singlet-extended SM due to theoretical considerations.

First, let us discuss limits coming from tree-level perturbative unitarity (see \eg \cite{Dicus2005}).
Hereby, the basic idea is that unitarity of the $S$-matrix constrains the theory's scattering amplitudes. In practice, this entails that the partial-wave amplitudes $a_j$ of a given process cannot be arbitrarily large. In the case of \textit{elastic} scattering of two \textit{identical} particles, for instance, the appropriate bound is
\begin{align}
	\absVal{\operatorname{Re} \tilde{a}_j(s)} \leq 1 \sepeq{} \forall j \geq 0 \fineq{,}
	\label{eq:SMS:unitarity}
\end{align}
which has to hold for all kinematically allowed center-of-mass energies $\sqrt{s}$.
The severest constraint usually originates from the $s$-wave amplitude, \ie $j=0$.
Importantly, the correct unitarity bound \eqref{eq:SMS:unitarity} contains modified partial-wave amplitudes $\tilde{a}_j$ to properly account for kinematical effects near threshold
\begin{align*}
	\tilde{a}_j(s) := \xi(s) a_j(s) \sepeq{\quad\text{with}}
	\xi(s) = \sqrt{1-\frac{4m^2}{s}} \fineq{.}
\end{align*}
The stated form of $\xi(s)$ is true for elastic scattering of two identical particles of mass $m$. As expected, we recover $\tilde{a}_j\to a_j$ for energies far above the threshold $s_\text{th}=4m^2$.

In order to further narrow down the model's viable parameter space, we now calculated the matrix elements of scalar two-to-two scattering processes. We identified the strongest constraints on the scalar couplings to come from light Higgs scattering, $hh\to hh$ (for calculational details, see Appendix \ref{app:unitarity}).
For one, near the kinematic threshold, contributions from virtual $h$-exchange dominate, in particular those from $t$- and $u$-channel diagrams.
In the $\theta\to 0$ (and hence $\Gamma_h\to 0$) limit, the corresponding expression for $\operatorname{Re}\tilde{a}_0$ exhibits a local extremum at $s/(4m_h^2) \simeq \num{1.4}$. Applying equation \eqref{eq:SMS:unitarity} then puts an upper limit on the trilinear coupling $\kappa_{3h}$:
Numerical evaluation yields
\begin{align}
	\absVal{\kappa_{3h}} \lesssim \num{1.64} \cdot m_h \quad
	\sepeq{\stackrel{\eqref{eq:SMS:smallThetaExp}}{\Longrightarrow}}
	\absVal{\kappa_3} \lesssim \num{4.9} \cdot m_h \fineq{.}
	\label{eq:SMS:unitarity_kappa3}
\end{align}
The above constraint confirms the intuitive expectation that a trilinear scalar self-coupling should not be much larger than the associated particle's mass (\cf also \cite{Schuessler2007}).
Let us furthermore remark that in the presence of a sizable quartic coupling $\lambda_{4h}$, the bound in equation \eqref{eq:SMS:unitarity_kappa3} is slightly relaxed due to cancellations between contributions from $h$-exchange diagrams and the contact interaction graph proportional to $\lambda_{4h}$.
For instance, the refined limit for $\kappa_4=2$ is $\absVal{\kappa_3/m_h} \lesssim \num{5.2}$.

A similar bound on $\delta_2$ can be inferred from the same process, $hh\to hh$. In the vicinity of the Higgs pole, the scattering amplitude is dominated by $s$-channel $H$-exchange and can become large. Hence, $\operatorname{Re}a_0$ possesses another local extremum near $s\simeq m_H^2$.
Applying equation \eqref{eq:SMS:unitarity} in this energy region limits the trilinear portal coupling, namely $\absVal{\kappa_{H2h}} \leq \sqrt{8\pi m_H \Gamma_H}$. For small scalar mixing angle, this translates to
\begin{align}
	\absVal{\delta_2} \leq \sqrt{32\pi m_H \Gamma_H^\text{SM}/v^2} \simeq \num{0.029} \fineq{,}
	\label{eq:SMS:unitarity_delta2}
\end{align}
where we used the SM prediction for the Higgs width \cite{DeFlorianSabaris:2215893} in evaluating equation \eqref{eq:SMS:unitarity_delta2}.
This is justified, since even adding the maximally allowed value from nonstandard Higgs decays, see equation \eqref{eq:SMS:GammaNonMax}, hardly alters the numerical result.
Note that this limit on $\delta_2$ is of the same order of magnitude yet slightly weaker than the one derived from nonstandard Higgs decays in the previous paragraph, see equation \eqref{eq:SMS:invisibleHiggs_delta2}.

Lastly, for asymptotically high energies, $s\to\infty$, only the contribution to $a_0(hh\to hh)$ due to momentum-independent contact interactions stays finite. By virtue of equation \eqref{eq:SMS:unitarity} this implies an upper bound on $\lambda_{4h}$, which then gives
\begin{align}
	\kappa_4 \leq \frac{8\pi}{3} \simeq \num{8.4}\fineq{}
	\label{eq:SMS:unitarity_kappa4}
\end{align}
in the limit of small $\theta$.

The second class of theoretical constraints emerges from the requirement of perturbativity of couplings. In other words, model parameters are to be chosen in such a way as to justify the perturbative expansion of physical observables.
For a generic dimensionless scalar coupling $\lambda$, one typically checks $\absVal{\lambda} \leq 4\pi$ to ensure validity of the perturbative expansion. However, the aforementioned upper bound is only applicable to a coupling which is normalized such that the associated Feynman rule is just $-\I \lambda$ without any numerical prefactors. Precisely in this case, the relevant expansion parameter of the perturbative series is $\lambda/(4\pi)$ and the above bound is meaningful.
Considering the low-energy Lagrangian in equation \eqref{eq:decays:potential}, we have to rescale the perturbativity limit accordingly
\begin{align}
	\lambda_{4h} \leq \frac{\pi}{6} \quad
	\sepeq{\Rightarrow} \kappa_4 \leq \frac{2\pi}{3} \simeq \num{2.1} \fineq{.}
	\label{eq:SMS:perturbativity_kappa4}
\end{align}
Compared to \eqref{eq:SMS:unitarity_kappa4}, this constraint on $\kappa_4$ is 4 times stronger.

A similar perturbativity bound can be derived for the quartic portal coupling $\lambda_{2H2h}$, which then translates to $\absVal{\delta_2} \leq 4\pi$ for small $\theta$. Obviously, this is not competitive to the limits on $\absVal{\delta_2}$ coming from experiment or unitarity.

%%%%%%%%%%%%%%%%%%%%%%%%%%%%%%%%%%%%%%
%%%%%%%%  S E C T I O N :  Phenomenological Consequences %%%%
%%%%%%%%%%%%%%%%%%%%%%%%%%%%%%%%%%%%%%

\section{Phenomenological consequences}
\label{sec:pheno}
\noindent
%================================================
Taking into account all of the constraints discussed above, we will now analyze the actual detection potential for scalar three-body Higgs decays, thereby pursuing the following argument:
For nonzero mixing, the light scalar $h$ will eventually decay to SM particles. The allowed final states thereby depend on its mass $m_h$.
Furthermore, as long as the mixing angle is not too small, the light scalar's total width $\Gamma_h$ is large enough so that its decays can take place within a typical detector radius.
Combined with a possibly sizable three-body scalar Higgs decay rate, this may give rise to unique signatures at current or future colliders.
To be able to calculate the expected cross sections of such characteristic processes in Section \ref{sec:pheno:3BH}, we need to first discuss the most important decays of $h$ in Section \ref{sec:pheno:hdecay}.
A complete phenomenological study including a full background analysis and dedicated Monte Carlo simulations is postponed to future work.

%%%%%%%%  S U B S E C T I O N :  Decays of the light scalar %%%%%%%%

\subsection{Decays of the light scalar}
\label{sec:pheno:hdecay}
\noindent
%================================================
Let us begin by considering the detection prospects for scalar three-body Higgs decays at a hadron collider like the LHC.
Here, final states containing muons seem to be particularly promising since all other particles that $h$ might decay into are harder to detect.%
\footnote{This is not true for electrons and photons. However, in the mass range of not too small scalar mixing, $m_h>\SI{360}{MeV}$, the corresponding branching ratios are already negligible.}
However, for the corresponding cross sections to be sufficiently large, the branching fraction $\Bmumu \equiv \mathcal{B}(h\to\mu^+\mu^-)$ must be sizable.
Requiring $m_h>\SI{360}{MeV}$ as argued in Section \ref{sec:SMS:constr}, the largest values for $\Bmumu$ can be achieved for masses below the kaon threshold $2m_K \approx \SI{988}{MeV}$.
The light scalar then predominantly decays into pions and muons.

Whereas the tree-level partial decay width into muons is straightforward to calculate,
\begin{align}
	\Gamma(h\to\mu^+\mu^-) = \sin^2\theta \frac{m_\mu^2 m_h}{8\pi v^2}
	\left( 1-\frac{4 m_\mu^2}{m_h^2} \right)^{3/2} \fineq{,}
	\label{eq:pheno:partialMuon}
\end{align}
the computation of $\Gamma(h\to\pi\pi)$ is more involved.\footnote{In the following, $\pi\pi$ refers to the sum of both contributing final states, \ie $\pi^+\pi^-$ and $\pi^0\pi^0$.} First, since $m_h$ is of the order of the QCD scale, perturbative methods are no longer applicable, and one has to resort to nonperturbative alternatives such as chiral perturbation theory ($\chi$PT). The partial width to leading order $\chi$PT is given by \cite{Voloshin1980,Voloshin1986,Chivukula:1989ds}
\begin{align*}
	\Gamma(h\to\pi\pi) = \sin^2\theta \frac{m_h^3}{216\pi v^2} \left( 1+\frac{11}{2}\frac{m_\pi^2}{m_h^2} \right)^2 \sqrt{1-\frac{4m_\pi^2}{m_h^2}} \fineq{.}
\end{align*}
Additionally, the above result will be modified by sizable effects due to final-state pion-pion interactions as first discussed in \cite{Raby1988b}. 
Note that all errors in $\Gamma(h\to\pi\pi)$ directly translate to uncertainties in the total decay width and thus also affect the branching fractions. Via equation \eqref{eq:SMS:upperLimitTheta} the upper limit on the scalar mixing angle in this mass region is sensitive to  the aforementioned uncertainties, as well.
In the following, we will use the results from \cite{Donoghue1990}, whose calculation is based on next-to-leading order $\chi$PT combined with dispersion theory to consistently account for final-state interactions. Their values of the ratio $b:=\Gamma(h\to\pi\pi)/\Gamma(h\to\mu^+\mu^-)$ are listed in Table \ref{tab:pheno:BRmumu} for various masses of the light scalar $h$. The observed large enhancement of $b$ for $m_h \gtrsim \SI{900}{MeV}$ is mainly due to the presence of the scalar isosinglet resonance $f_0(980)$. Given $b$, the total width of $h$ and its branching fraction into muons are given by
\begin{align}
	\Gamma_h = (1+b) \Gamma(h\to\mu^+\mu^-) \sepeq{\;\;\text{and}} 
	\Bmumu = \frac{1}{1+b} \fineq{.}
	\label{eq:pheno:ObservablesFromRatio}
\end{align}

\begin{table}[t]
	\centering
	\begin{tabular}{llccccccc}
		\toprule
		$m_h$ & [GeV] & \symhspace{0.5em}{0.4} & \symhspace{0.5em}{0.5} & \symhspace{0.5em}{0.6} & \symhspace{0.5em}{0.7} & \symhspace{0.5em}{0.8} & \symhspace{0.5em}{0.9} & \symhspace{0.5em}{0.95} \\
		\colrule
		\vspace{-0.9em}\\
		$b$ & & 6 & 10 & 12 & 16 & 21 & 44 & 91 \\
		$\Bmumu$ & [\%] & 14.3 & 9.1 & 7.7 & 5.9 & 4.5 & 2.2 & 1.1 \\
		$\sin\theta_\text{max}$ & $[10^{-3}]$ & 1.9 & 2.4 & 2.6 & 2.9 & 3.4 & 4.8 & 6.8 \\
		\botrule
	\end{tabular}
	\caption{The ratio $b\equiv \Gamma_{\pi\pi}/\Gamma_{\mu\mu}$ is obtained from \cite{Donoghue1990}. The branching fraction into muons then follows from equation \eqref{eq:pheno:ObservablesFromRatio}. Finally, the experimental upper limit on the scalar mixing angle $\theta$ is inferred from equation \eqref{eq:SMS:upperLimitTheta}.}
	\label{tab:pheno:BRmumu}
\end{table}

\begin{figure}[b]
	\hspace{-0.3em}
	\includegraphics[scale=0.757]{\figPath/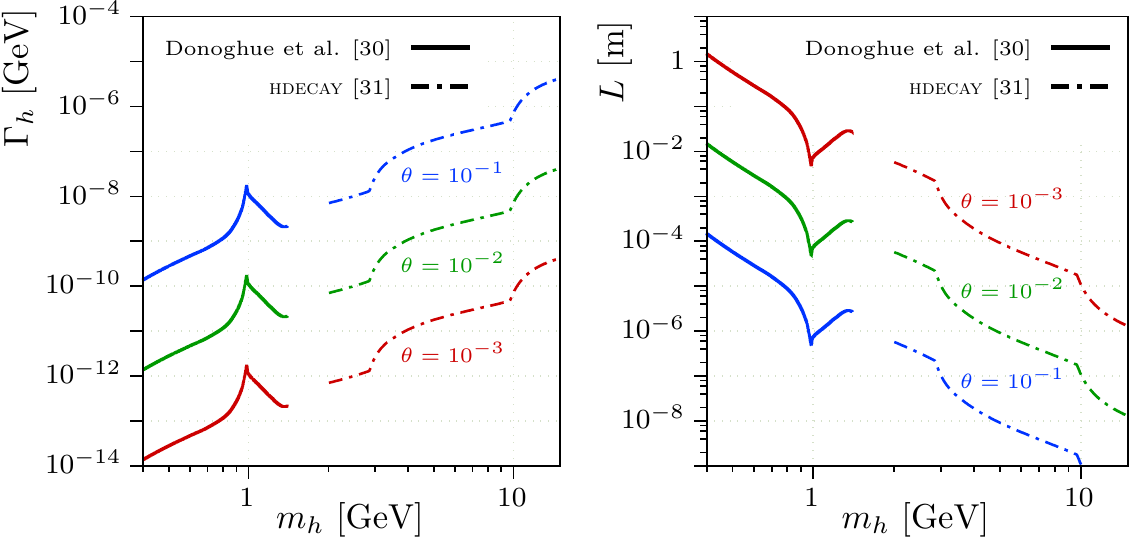}	% previously scale=0.78
	\caption{Light scalar's mass-dependent decay width (\textit{left}) and decay length (\textit{right}) for various scalar mixing angles $\theta$. The total width was calculated using the results of \cite{Donoghue1990} ($m_h\leq\SI{1.4}{GeV}$) and \cite{Djouadi1998} ($m_h\geq\SI{2}{GeV}$), respectively. The decay length was obtained via equation \eqref{eq:pheno:decayLen} assuming $E_h\simeq m_H/3$.}
	\label{fig:pheno:decayLen}
\end{figure}

As we consider larger masses of up to \SI{10}{GeV}, more and more decay channels open such that $\Bmumu$ significantly decreases, and the muon final state becomes irrelevant.
At the same time, for $m_h\gtrsim\SI{1}{GeV}$ nonperturbative QCD effects become less important and the branching ratios of $h$ in this regime tend to those of a SM Higgs boson with the same mass, which may then be easily calculated using appropriate tools like \textsc{hdecay} \cite{Djouadi1998}.
The dominant decays are now into pairs of tau leptons, gluons, $c$- and $b$-quarks.
While detecting scalar three-body Higgs decays via tau final states may well be feasible at the LHC, searching for the other channels will be difficult.
Nevertheless, it is still interesting to know the corresponding rates in view of future $e^+ e^-$ colliders like the ILC.

Finally, let us very briefly comment on the light scalar's typical decay length $L$.
In the lab frame, $h$ will be produced with an energy of order of the Higgs mass, \ie $E_h = \mathcal{O}(m_H) \gg m_h$. In this limit, the decay length can be computed as
\begin{align}
	L = \frac{E_h}{m_h \Gamma_h} + \mathcal{O}\left(\frac{m_h^2}{E_h^2}\right) \fineq{.}
	\label{eq:pheno:decayLen}
\end{align}
The largest values for $L$ are obtained in the low-mass regime, where there is only a small number of open final states and the mixing angle is necessarily tiny.
For instance, for a \SI{500}{MeV} scalar with energy $E_h \simeq m_H/3$, one finds using equations \eqref{eq:pheno:partialMuon} to \eqref{eq:pheno:decayLen}
\begin{align}
	L \simeq \left( \frac{10^{-3}}{\sin\theta} \right)^{\!2} \cdot \SI{0.55}{m} \fineq{,}
	\label{eq:pheno:decayLenNum-500MeV}
\end{align}
demonstrating that we expect light scalar decays to happen at clearly displaced vertices.
In the high-$m_h$ regime the light scalar's total width increases significantly. Employing \textsc{hdecay}, we exemplarily calculate the decay length for $m_h=\SI{5}{GeV}$ and $E_h \simeq m_H/3$
\begin{align}
	L \simeq \left( \frac{0.1}{\sin\theta} \right)^{\!2} \cdot \SI{9.1}{nm} \fineq{,}
	\label{eq:pheno:decayLenNum-5GeV}
\end{align}
thus showing that displaced vertices are not a feature for larger masses of $h$ in the case of typical values for $\sin\theta$.
A more thorough overview of the typical light scalar's decay length is given in Figure \ref{fig:pheno:decayLen}.

%%%%%%%%  S U B S E C T I O N :  Scalar Higgs decays at colliders %%%%%%%%

\subsection{Scalar Higgs decays at colliders}
\label{sec:pheno:3BH}
\noindent
%================================================
In the following, we will apply the formulas for the scalar Higgs decays from Section \ref{sec:decays} to the singlet-extended SM introduced above.
Thereby, we will take into account all the constraints on the model's couplings discussed in the previous paragraphs.
We will first focus on the question of whether there are valid points in parameter space for which the three-body rate becomes comparable to that of two-body decays.
Afterwards, we will calculate the typical cross sections of processes that might lead to discovering scalar three-body Higgs decays at a collider experiment.

%%%%%%%%  S U B S U B S E C T I O N :  Low-mass regime %%%%%%%%

\subsubsection{Low-mass regime}
\begin{figure}[t]
	\centering
	\includegraphics[scale=0.79]{\figPath/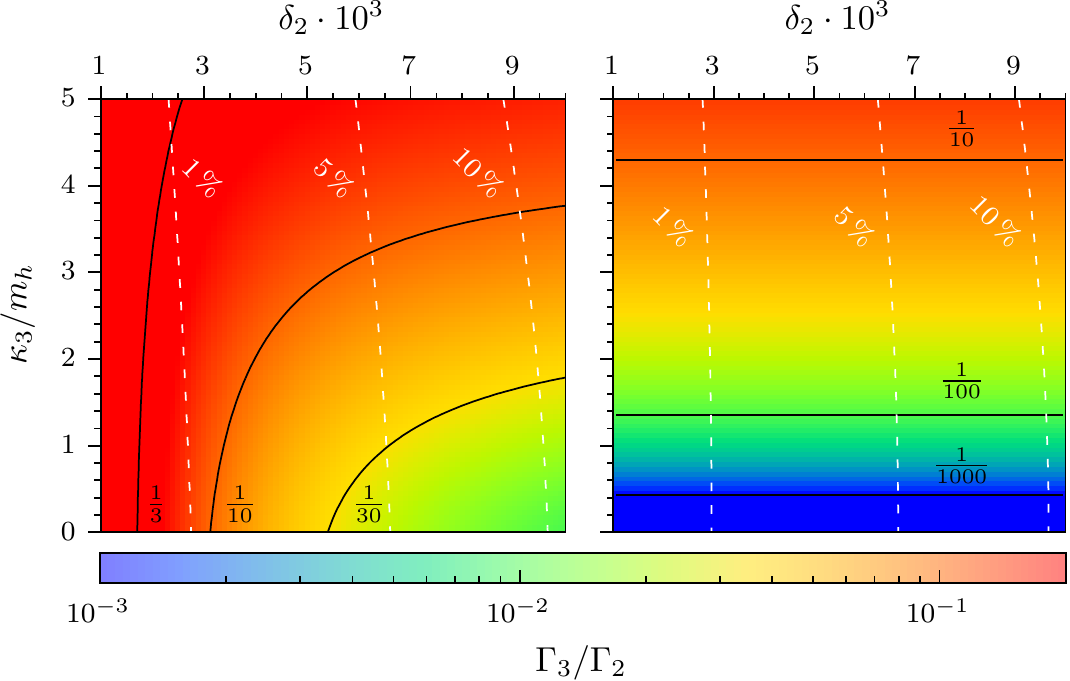}
	\caption{Ratio $r=\Gamma_3/\Gamma_2$ of scalar three-body and two-body Higgs decay rates. The light scalar mass was fixed to $m_h=\SI{0.9}{GeV}$. Black solid lines are the contours of constant $r$. Additionally, we show contours of constant nonstandard Higgs branching fraction $\mathcal{B}_\text{non}=(\Gamma_2+\Gamma_3)/\Gamma_H$ as white dashed lines. \textit{Left}: Large mixing ($\sin\theta=\num{4.8e-3}$) and self-coupling ($\kappa_4=\num{2.0}$). \textit{Right}: Small mixing ($\sin\theta \to 0$) and self-coupling ($\kappa_4=\num{0.1}$).}
	\label{fig:pheno:2Dscan-900MeV}
\end{figure}

\noindent
%================================================
Let us begin our study in the low-mass regime, where the light scalar lies below the kaon threshold $2m_K$.
Here, we first determine how the ratio $r = \Gamma_3/\Gamma_2$ introduced in equation \eqref{eq:decays:ratio} depends on the model parameters.
For that purpose, we show the results of two-dimensional parameter scans in the $\delta_2$-$\kappa_3$ plane in Figure \ref{fig:pheno:2Dscan-900MeV}.
Here, the color code represents the size of $r$ with the associated contours drawn in black. Additionally, we show contours of constant nonstandard Higgs branching fraction $\mathcal{B}_\text{non} \equiv (\Gamma_2+\Gamma_3)/\Gamma_H$ as white dashed lines.
The left and right panel in Figure \ref{fig:pheno:2Dscan-900MeV} differ in the used values for mixing angle $\theta$ and quartic singlet self-interaction $\kappa_4$. Physically, these two parameters determine the relevance of the contribution $\Gamma_3^c$ of contact interactions to the three-body width, which is hence relatively large (small) in the left (right) image.
In both scans, we set the light scalar's mass to $m_h = \SI{900}{MeV}$ which is near but still below the kaon threshold.
Note that, as long as $m_h$ is much smaller than the LHC Higgs mass, its direct effect on $r$ is negligible. However, since the exact value of $m_h$ crucially influences the maximally allowed mixing angle via Table \ref{tab:pheno:BRmumu}, $r$ can still indirectly depend on $m_h$.

As demonstrated by the nearly vertical white contours in both scenarios of Figure \ref{fig:pheno:2Dscan-900MeV}, the size of the nonstandard Higgs decay width is mainly determined by $\delta_2$ as expected for tiny $\theta$.
Also the role of $\kappa_3$ as anticipated in Section \ref{sec:SMS} is confirmed here: the larger $\kappa_3$ the more important the three-body final state's relative contribution to nonstandard Higgs decays.
In particular, even for vanishing mixing and small $\kappa_4$, one obtains ratios as large as $\mathcal{O}(\num{0.1})$ provided $\kappa_3$ is sufficiently big (\cf right panel).
Note, however, that the above rule does not properly describe the leftmost part of the left panel, where $r$ is large and almost constant over the entire $\kappa_3$ range.
The reason is that both $\Gamma_2$ and $\Gamma_3^h$ \textendash\ the contribution to $\Gamma_3$ due to $h$-exchange \textendash\ decrease with $\delta_2$, whereas contact interactions, $\Gamma_3^c$, stay constant.
Most importantly, this behavior allows the ratio $r$ to become of order one for small enough $\delta_2$. We thus demonstrated that one can indeed find regions of parameter space, where two- and three-body Higgs decays are equally relevant.
This is, however, not possible in the absence of contact interactions, where two- and three-body rates drop in equal measure with $\delta_2$ so that $r$ is constant in $\delta_2$ (right panel).

Revisiting our discussion on requirements for large $r$ from the end of Section \ref{sec:decays}, let us stress a further crucial point here:
Provided only $h$-mediated diagrams contribute to $\Gamma_3$, we see from the right panel of Figure \ref{fig:pheno:2Dscan-900MeV} that $r$ can maximally become of order \num{0.1} due to a bound on $\kappa_{3h}$ from perturbative unitarity.
To appreciate the necessity of this limitation assume, for the moment, that significantly larger values for $\kappa_{3h}$ were allowed.
Then, on the one hand, $r\simeq 1$ could be realized even without contact interactions.
On the other hand, equally large rates for scalar $n$-body Higgs decays with $n\ge 4$ might become possible.
These would arise from $h$-exchange diagrams similar to that in Figure \ref{fig:decays:3BodyTree} but with more internal scalar propagators and/or additional four-point vertices.
The resulting scenario with $\Gamma_2 \simeq \Gamma_3 \simeq \Gamma_4 \simeq \ldots$ would be clearly unphysical.
Crucially, elevating $r$ from $\mathcal{O}(\num{0.1})$ to $\mathcal{O}(1)$ or larger must therefore inevitably come from additional tree-level contact interactions (as in Figure \ref{fig:pheno:2Dscan-900MeV}, left panel).
In an effective theory, those interactions for $n\ge 4$ correspond to irrelevant operators and are thus typically suppressed by some high mass scale. In a renormalizable theory, irrelevant operators are entirely forbidden. Hence, three-body decays are indeed a special case.

\begin{table}[b]
	\centering
	\begin{tabular}{l|cccc}
	\toprule
	Point & \symhspace{1.0em}{$\delta_2$} & \symhspace{1.0em}{$\Gamma_2$ [GeV]} & \symhspace{1.0em}{$\Gamma_3$ [GeV]} & \symhspace{1.0em}{$\Gamma_4$ [GeV]} \\
	\colrule
	\vspace{-0.9em}\\
	\enquote{best} & $10^{-3}$ & \num{5.3e-6} & \num{1.1e-6} & \num{2.0e-7} \\
	\hspace{0.25em}\eqref{eq:pheno:bestPt} & $10^{-2}$ & \num{4.9e-4} & \num{6.4e-5} & \num{8.8e-6} \\
	\colrule
	\vspace{-0.9em}\\
	\enquote{worst} & $10^{-3}$ & \num{4.8e-6} & \num{2.6e-8} & \num{2.8e-10} \\
	\hspace{0.25em}\eqref{eq:pheno:worstPt} & $10^{-2}$ & \num{4.8e-4} & \num{2.6e-6} & \num{2.8e-8} \\
	\botrule
	\end{tabular}
	\caption{Higgs decay rates $\Gamma_n$ into $n$ light scalars of mass $m_h=\SI{500}{MeV}$ for the \enquote{best-case} and \enquote{worst-case} parameter points from equations \eqref{eq:pheno:bestPt} and \eqref{eq:pheno:worstPt}, respectively.}
	\label{tab:pheno:rates}
\end{table}

Now that we have seen how the ratio $r$ depends on the model parameters, it will be interesting to study specific observables which might help to \textit{directly} measure three-body Higgs decays at collider experiments as the LHC.
In doing so, we will distinguish two cases, generically denoted as \enquote{best} and \enquote{worst} case, respectively.
On the one hand, in the \enquote{best-case} scenario, we will assume that all model parameters saturate their respective bounds (\cf Section \ref{sec:SMS:constr} and Table \ref{tab:pheno:BRmumu}). Physically, this means that the singlet sector couples reasonably strong to the SM and exhibits relatively strong self-interactions.
On the other hand, the decoupling limit, $\theta\to 0$, with the trilinear singlet coupling attaining its natural value, \ie $\kappa_3 \simeq m_h$, constitutes the \enquote{worst-case} scenario.
To be more specific, we will now consider the following benchmark points, fixing $m_h=\SI{500}{MeV}$
\begin{subequations}
	\begin{align}
		\label{eq:pheno:bestPt}
		\kappa_3 & =\SI{2.45}{GeV} \sepeq{,} \kappa_4 = \num{1.0} \sepeq{,} \sin\theta = \num{0.0024} \\
		\label{eq:pheno:worstPt}
		\kappa_3 & =\SI{0.50}{GeV} \sepeq{,}\hspace{0.02em} \kappa_4 = \num{0.1} \sepeq{,} \sin\theta = 0.0
	\end{align}
	\label{eq:pheno:benchmark}%
\end{subequations}%
Table \ref{tab:pheno:rates} contains the corresponding Higgs decay rates to two, three and four light scalars.%
\footnote{The \textit{four}-body decay rates $\Gamma_4$ were calculated using \textsc{CalcHEP} \cite{Belyaev2013} in order to demonstrate that the expected hierarchy of partial widths is intact, $\Gamma_3 \gg \Gamma_4$.}
For each point, we consider two different values of the portal coupling $\delta_2$ in order to demonstrate the rates' overall scaling (\cf white, dashed contour lines in Figure \ref{fig:pheno:2Dscan-900MeV}).

As we have argued in Section \ref{sec:pheno:hdecay}, a \SI{500}{MeV} Higgs-like scalar will predominantly decay to muon or pion pairs, respectively.
This shows that scalar three-body Higgs decays may lead to very special final states, the most spectacular of which would be one containing six muons, \ie
\begin{align*}
	pp \to H \to 3 h \to 6 \mu \fineq{.}
\end{align*}
Considering that $m_H\gg m_h$, the light scalars will be strongly boosted. Thus one actually expects three pairs of collimated muons each having the same invariant mass $m_h$.
Furthermore, sizable light scalar lifetimes imply that the muons start from secondary vertices (\cf equation \eqref{eq:pheno:decayLenNum-500MeV}).
Obviously, this is a very clean signature with small systematic uncertainties and little background.
But also final states where one or more muon pairs are replaced by pions may be interesting.
Although they are not as clean at the LHC as the purely leptonic channel, they provide larger rates since $h$ decays mainly into pions.

In order to assess the prospects for actually observing these processes at the LHC, we compute their expected cross sections for the benchmark points in \eqref{eq:pheno:benchmark}. Applying the narrow width approximation twice, one obtains
\begin{align}
	\sigma_{2n\mu} :=
	\sigma_\text{prod} \cdot
	\mathcal{B}(H\to n h) \cdot
	\mathcal{B}^n(h\to \mu^+\mu^-) \fineq{.}
	\label{eq:pheno:cross-section}
\end{align}
Formulas of similar form hold for the various other final states mentioned above.%
\footnote{Note that an extra combinatorial factor $k$ has to be included for processes where not all light scalars decay to the same final state. In case of the $4\mu2\pi$ final state, for instance, one has $k=3$.}
A few comments on equation \eqref{eq:pheno:cross-section} are in order. First, we denote the production cross section of the heavy Higgs by $\sigma_\text{prod}$. It is given by multiplying the corresponding SM value by $\cos^2\theta \approx 1$.
Higgs production at the LHC is dominated by gluon fusion, the cross section of which at \SI{14}{TeV} for a \SI{125}{GeV} SM Higgs boson is roughly $\sigma_\text{prod} \approx \SI{55}{pb}$ \cite{DeFlorianSabaris:2215893}.
Secondly, in order to compute the heavy Higgs branching fractions we employ a properly adapted total Higgs decay width, $\Gamma_H \approx \cos^2\theta\cdot\Gamma_H^\text{SM} + \Gamma_2 + \Gamma_3$, where we assumed all decay rates $\Gamma_n$ to be negligible for $n\geq4$.
Lastly, the light scalar's branching ratios were taken from Table \ref{tab:pheno:BRmumu}.

\begin{table}[t]
	\centering
	\begin{tabular}{l|ccccc}
	\toprule
	Point & \symhspace{1.0em}{$\delta_2$} & \symhspace{0.6em}{$\sigma_{4\mu}$ [fb]} & \symhspace{0.4em}{$\sigma_{6\mu}$ [fb]} & \symhspace{0.2em}{$\sigma_{4\mu2\pi}$ [fb]} & \symhspace{0.2em}{$\sigma_{2\mu4\pi}$ [fb]} \\
	\colrule
	\vspace{-0.9em}\\
	\enquote{best} & $10^{-3}$ & \num{0.59} & \num{1.1e-2} & \num{0.33} & \num{3.3} \\
	\hspace{0.25em}\eqref{eq:pheno:bestPt} & $10^{-2}$ & 48 & \num{0.57} & 17 & \num{1.7e2} \\
	\colrule
	\vspace{-0.9em}\\
	\enquote{worst} & $10^{-3}$ & \num{0.54} & \num{2.6e-4} & \num{7.8e-3} & \num{7.8e-2} \\
	\hspace{0.25em}\eqref{eq:pheno:worstPt} & $10^{-2}$ & 48 & \num{2.3e-2} & \num{0.70} & \num{7.0} \\
	\botrule
	\end{tabular}
	\caption{Total cross sections of characteristic final states in $\sqrt{s}=\SI{14}{TeV}$ $pp$ collisions at the LHC. Higgs production via gluon fusion is assumed.}
	\label{tab:pheno:cross-sections-LHC}
\end{table}

Our numerical results for different final states containing at least one muon pair are listed in Table \ref{tab:pheno:cross-sections-LHC}.
With an assumed integrated luminosity of $\SI{300}{fb^{-1}}$ (or $\SI{3000}{fb^{-1}}$ after the planned luminosity upgrade), the six-muon final state is unlikely to be seen in the \enquote{worst-case} scenario at the LHC.
However, for large enough couplings as for the benchmark point \eqref{eq:pheno:bestPt}, it might be in reach of the \SI{14}{TeV} run where one expects up to $\mathcal{O}(100)$ six-muon events. In the high-luminosity phase, this number increases to $\mathcal{O}(1000)$.
The rates are even higher by up to 2 orders of magnitude if final-state muon pairs are exchanged for pions. These channels might even probe scenarios with smaller couplings as realized in benchmark point \eqref{eq:pheno:worstPt}. Here, one expects up to $\mathcal{O}(100)$ events in the $4\mu2\pi$ channel and $\mathcal{O}(1000)$ in the $2\mu4\pi$ channel with $\SI{300}{fb^{-1}}$ of data.
Note that, since the cross sections for the final states containing muons are proportional to an appropriate power of $\Bmumu$, they become smaller (larger) with growing (decreasing) mass $m_h$ (see Table \ref{tab:pheno:BRmumu}).

Although the above numbers look already very promising, it is important to study the aforementioned processes in more detail in the context of dedicated event and detector simulations
including potential backgrounds and various detector efficiencies.%
\footnote{For instance, collimated muon and pion pairs from displaced vertices are challenging for both trigger and reconstruction \cite{ATL-PHYS-PUB-2016-010,ATLAS:2016jza}.}
Here, we only briefly mention that the main SM background to the $6\mu$ final state will probably come from associated $t\bar{t}W$, $t\bar{t}Z$ and $t\bar{t}b\bar{b}$ production with subsequent (semi)leptonic decays of the particles involved.
However, the corresponding cross sections are rather small \cite{Frixione2015,Bredenstein2009}.
Besides, muons from $b$ decays can be efficiently rejected by imposing appropriate dimuon isolation cuts (see \eg \cite{Khachatryan2016}).
Requiring the invariant masses of all dimuons to be compatible will further reduce background.
A detailed study of this subject is postponed to future work.

As demonstrated above, the more pions a six-particle final state contains the larger is the process's rate (Table \ref{tab:pheno:cross-sections-LHC}).
The six-pion final state, however, is very challenging to search for at the LHC due to, among other things, the large amount of QCD background.
In contrast, it might be the most promising channel to observe scalar three-body Higgs decays in the low-$m_h$ regime at an $e^+ e^-$ collider like the ILC.
In Table \ref{tab:pheno:cross-sections-ILC} we therefore list the corresponding cross sections at the ILC operating above the $ZH$ production threshold at $\sqrt{s}\simeq\SI{250}{GeV}$ with polarized beams. The cross section for the \SI{125}{GeV} Higgs to be produced via the Higgs-strahlung channel is then $\sigma_\text{prod}\approx\SI{300}{fb}$ \cite{Baer2013}.
During its \SI{250}{GeV} run the ILC is planned to collect an integrated luminosity of $\SI{500}{fb^{-1}}$ (or $\SI{1500}{fb^{-1}}$ after a luminosity upgrade) \cite{Barklow2015}.
Thus, Table \ref{tab:pheno:cross-sections-ILC} demonstrates that the six-muon channel is very unlikely to be seen at the ILC.
In contrast, we expect up to $\mathcal{O}(1000)$ three-body scalar Higgs decays with a six-pion final state in the \enquote{best-case} scenario.
But even for the conservative benchmark point from equation \eqref{eq:pheno:worstPt}, up to $\mathcal{O}(100)$ six-pion events from three-body scalar Higgs decays are possible, and an observation may be feasible.
Note finally that for increasing $m_h$, the pion channel is expected to be even more abundant since the branching fraction for a decay of $h$ into pions grows with $m_h$ (\cf Table \ref{tab:pheno:BRmumu}).

\begin{table}[t]
	\centering
	\begin{tabular}{l|ccccc}
	\toprule
	Point & \symhspace{1.0em}{$\delta_2$} & \symhspace{0.6em}{$\sigma_{4\mu}$ [fb]} & \symhspace{0.6em}{$\sigma_{6\mu}$ [fb]} & \symhspace{0.6em}{$\sigma_{4\pi}$ [fb]} & \symhspace{0.6em}{$\sigma_{6\pi}$ [fb]} \\
	\colrule
	\vspace{-0.9em}\\
	\enquote{best} & $10^{-3}$ & \num{3.2e-3} & \num{6.1e-5} & \num{0.32} & \num{6.1e-2} \\
	\hspace{0.25em}\eqref{eq:pheno:bestPt} & $10^{-2}$ & \num{0.26} & \num{3.1e-3} & 26 & \num{3.1} \\
	\colrule
	\vspace{-0.9em}\\
	\enquote{worst} & $10^{-3}$ & \num{2.9e-3} & \num{1.4e-6} & \num{0.29} & \num{1.4e-3} \\
	\hspace{0.25em}\eqref{eq:pheno:worstPt} & $10^{-2}$ &  \num{0.26} & \num{1.3e-4} & 26 & \num{0.13} \\
	\botrule
	\end{tabular}
	\caption{Total cross sections of characteristic final states in $\sqrt{s}=\SI{250}{GeV}$ $e^+e^-$ collisions at the ILC. Higgs production via Higgs-strahlung is assumed.}
	\label{tab:pheno:cross-sections-ILC}
\end{table}

%%%%%%%%  S U B S U B S E C T I O N :  Intermediate-mass regime %%%%%%%%

\subsubsection{Intermediate-mass regime}
\noindent
%================================================
Let us now analyze larger masses for $h$. Note, however, that the intermediate-mass regime between the kaon and the $B$-meson threshold at $m_B\approx\SI{5}{GeV}$ is not particularly promising for LHC searches for three-body scalar Higgs decays.
For one, the mixing angle $\theta$ remains tightly constrained by $B$-meson decay measurements.
At the same time, the light scalar's branching fraction into muons even further decreases due to the presence of an additional decay channel, $h\to \bar{K}K$.
Thus, the (semi)leptonic final states will have a tiny rate while the purely hadronic channels involving kaons and/or pions suffer from much QCD background as before.
In contrast, the hadronic final states may be observable in the much cleaner environment of an $e^+ e^-$ machine like the ILC. In Table \ref{tab:pheno:intermediateMassNumbers} we therefore list the corresponding event numbers for a \enquote{best-case} benchmark point assuming a light scalar mass of \SI{1.2}{GeV}.

\begin{table}[h]
	\centering
	\begin{tabular}{l|ccccc}
	\toprule
	\multirow{2}{*}{$\mathcal{B}$ [\%]} & \symhspace{0.25em}{$H\to 2h$} & \symhspace{0.25em}{$H\to 3h$} & \symhspace{0.25em}{$h\to\mu^+\mu^-$} & \symhspace{0.25em}{$h\to\pi\pi$} & \symhspace{0.25em}{$h\to\bar{K}K$} \\
	& \num{12.3} & \num{1.77} & \num{2.3} & \num{10.5} & \num{87.2} \\
	\colrule
	\multirow{2}{*}{ILC} & $N_{4K}$ & $N_{6K}$ & $N_{6\pi}$ & $N_{2\mu 4K}$ & $N_{2\pi 4K}$ \\
	& \num{14071} & 1760 & 3 & 139 & 636 \\
	\botrule
	\end{tabular}
	\caption{Branching ratios and event numbers for a $\SI{1.2}{GeV}$ singletlike Higgs at the \SI{250}{GeV} ILC run with $\SI{500}{fb^{-1}}$ of accumulated data. The remaining model parameters were set to $\sin\theta=\num{4.8e-3}$, $\kappa_3/m_h = \num{4.9}$, and $\delta_2 = \num{0.01}$, $\kappa_4 = \num{2.0}$.}
	\label{tab:pheno:intermediateMassNumbers}
\end{table}

%%%%%%%%  S U B S U B S E C T I O N :  High-mass regime %%%%%%%%

\subsubsection{High-mass regime}
\noindent
%================================================
As we have seen in Section \ref{sec:SMS:constr}, the bounds on $\theta$ relax significantly for light masses above the $B$-meson threshold, \ie $\SI{5}{GeV} \lesssim m_h \ll m_H$.
The light scalar will now predominantly decay into a pair of tau leptons or two jets which makes direct detection of three-body scalar Higgs decays at the LHC challenging also in this mass region.
However, the potentially larger mixing angles $\theta=\mathcal{O}(\num{0.1})$ has two important consequences that may qualitatively change the model's phenomenology.
On the one hand, the contact interaction diagram of Figure \ref{fig:decays:3BodyTree} can now give the dominant contribution to the scalar three-body Higgs decays if $\kappa_4$ is sufficiently large.
On the other hand, the effective trilinear portal $\kappa_{H2h}$ can now become anomalously small.
According to equation \eqref{eq:SMS:smallThetaExp} and assuming $\delta_2>0$, this happens near
\begin{align*}
	\kappa_{H2h} \simeq 0 \sepeq{\quad\Leftrightarrow\;} 
	\kappa_3 \simeq - \left( \frac{0.1}{\theta} \right) \left( \frac{\delta_2}{10^{-3}} \right) \cdot \SI{1.2}{GeV}
	\fineq{.}
\end{align*}
Here, the two-body rate $\Gamma_2$ tends to zero, whereas the three-body rate $\Gamma_3$ stays finite, leading to values for $r = \Gamma_3/\Gamma_2$ significantly larger than one.

\begin{figure}[b]
	\centering
	\includegraphics[scale=0.79]{\figPath/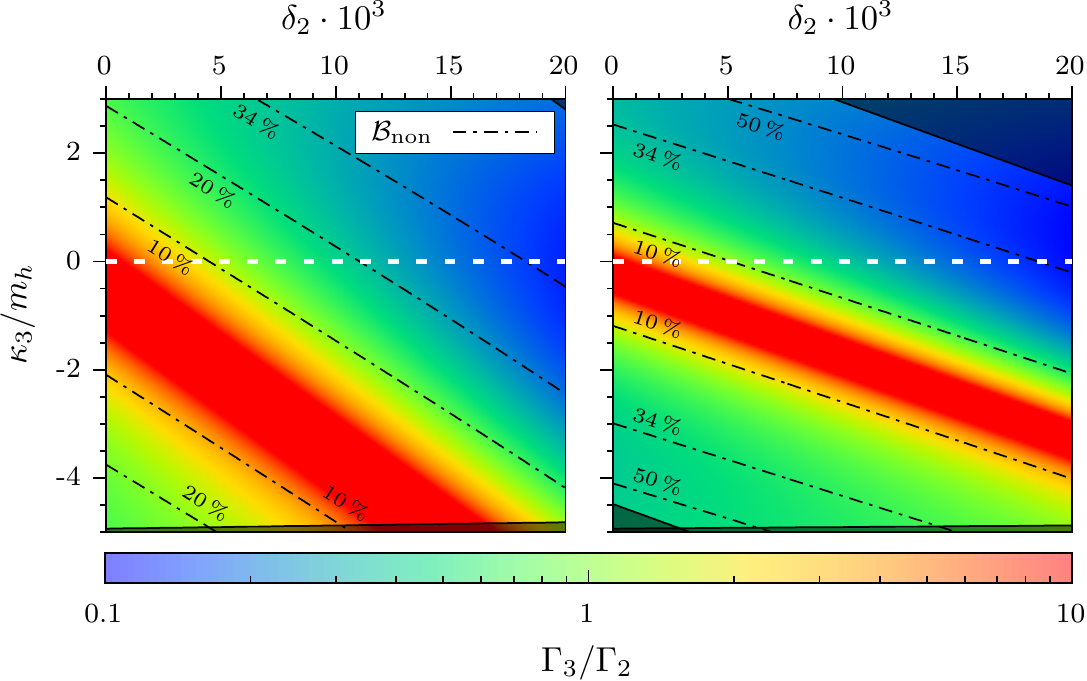}
	\caption{Ratio $\Gamma_3/\Gamma_2$ of scalar three-body and two-body Higgs decay rates.
	The light scalar mass was fixed to $m_h=\SI{5}{GeV}$ (\textit{left}) and \SI{10}{GeV} (\textit{right}), respectively.
	The remaining model parameters were set to $\sin\theta=\num{0.08}$ and $\kappa_4=\num{1.0}$.
	As black dash-dotted lines, we show contours of constant nonstandard Higgs branching fraction $\mathcal{B}_\text{non}$.
	Dark shaded regions are inconsistent with tree-level perturbative unitarity (\cf equations \eqref{eq:unitarity:bound-HiggsPole} and \eqref{eq:unitarity:bound-Threshold}).
	The model's phenomenology along the white dashed lines is investigated in more detail in Figure \ref{fig:pheno:cross-sections-5GeV}.}
	\label{fig:pheno:2Dscan-5GeV-10GeV}
\end{figure}

We show the results of a two-dimensional parameter scan in the $\delta_2$-$\kappa_3$ plane for a light scalar $h$ of mass $m_h = \SI{5}{GeV}$ (\SI{10}{GeV}) in the left (right) panel of Figure \ref{fig:pheno:2Dscan-5GeV-10GeV}.
Similar to Figure \ref{fig:pheno:2Dscan-900MeV}, the color encodes the size of $r$ while the contours are lines of constant nonstandard Higgs branching fraction $\mathcal{B}_\text{non}$.
Most importantly, note that $r$ is of order one or larger over a considerable part of the experimentally allowed region ($\mathcal{B}_\text{non}\leq \SI{34}{\%}$).
As an aside, we mention that null results of existing LHC searches for light bosons might further constrain the region \textit{outside} the red-yellow bands of Figure \ref{fig:pheno:2Dscan-5GeV-10GeV}, where the two-body rate becomes sizable (see \eg \cite{Aggleton2016} for a recent overview).
In contrast, the \textendash\ from our paper's point of view \textendash\ particularly interesting \textit{interior} of the red-yellow bands cannot be affected by the aforementioned searches, which is why we do not take them into account here.

\begin{figure}[t]
	\centering
	\includegraphics[scale=0.78]{\figPath/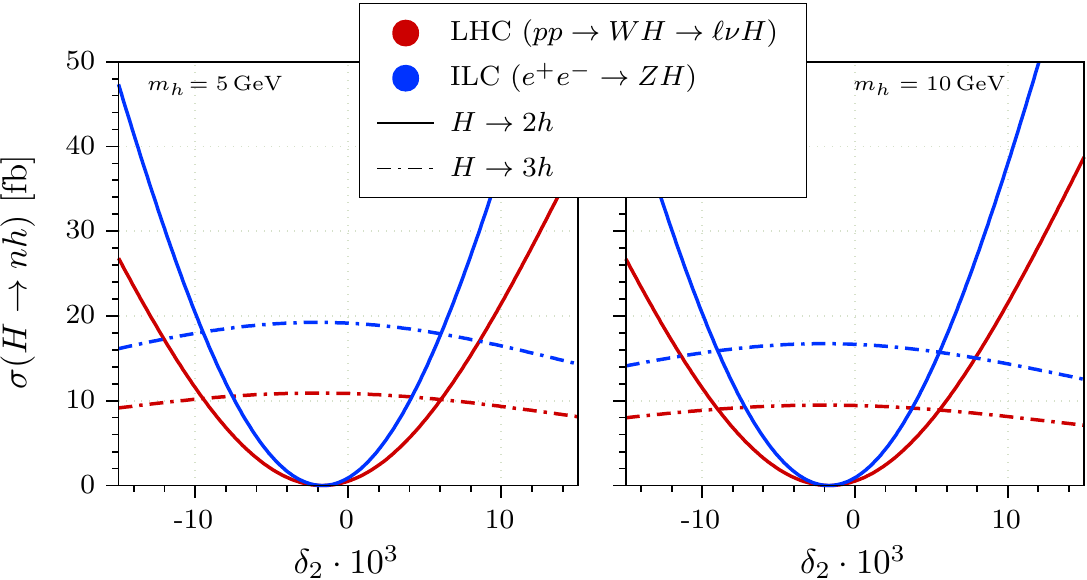}
	\caption{Cross section for scalar two- and three-body Higgs decays with $m_h=\SI{5}{GeV}$ (\textit{left}) and $m_h=\SI{10}{GeV}$ (\textit{right}) at the LHC (red) and ILC (blue), respectively. Model parameters are set to $\sin\theta=\num{0.08}$, $\kappa_{3}=0$ and $\kappa_4=\num{1.0}$. For the LHC, $W$-associated Higgs production with subsequent leptonic $W$ decay at \SI{14}{TeV} is assumed. The experimental bound on nonstandard Higgs decays, $\mathcal{B}_\text{non}\leq \SI{34}{\%}$ \cite{Aad2016z}, is satisfied over the whole range of the plots.}
	\label{fig:pheno:cross-sections-5GeV}
\end{figure}

Finally, in order to assess the detection prospects for scalar three-body Higgs decays in the high-$m_h$ regime, we exemplarily investigate the model's phenomenology in more detail for two different scalar masses and vanishing $\kappa_3$ (\cf dashed white lines in Figure \ref{fig:pheno:2Dscan-5GeV-10GeV}). 
Our findings are presented in Figure \ref{fig:pheno:cross-sections-5GeV}.
Here, we calculated the expected cross sections for two- and three-body scalar Higgs decays during the \SI{14}{TeV} LHC and the \SI{250}{GeV} ILC run.
Note that we did not specify how the light scalars will eventually decay. Accordingly, results from the figure must be multiplied by appropriate branching fractions of $h$ in order to obtain exclusive cross sections (\cf equation \eqref{eq:pheno:cross-section}).
Note furthermore that, unlike before, we considered $W$-associated Higgs production with subsequent leptonic $W$ decay for the LHC results.
The corresponding cross section for a \SI{125}{GeV} Higgs is $\sigma_\text{prod}\approx\SI{0.17}{pb}$ \cite{DeFlorianSabaris:2215893}.
For final states containing only jets otherwise, the presence of the extra lepton is crucial for triggering and background reduction.
However, if one or more light scalars decay into tau pairs, an analysis based on Higgs production via vector boson fusion (VBF) or gluon fusion (ggH) should be possible since there is a sizable probability of at least one lepton from tau decays.
In these cases the LHC cross sections of Figure \ref{fig:pheno:cross-sections-5GeV} increase by roughly 1 (VBF) or 2 (ggH) orders of magnitude.

Most importantly, Figure \ref{fig:pheno:cross-sections-5GeV} demonstrates that cross sections for scalar three-body Higgs decays are of order \SI{10}{fb} over the whole $\delta_2$ range for both colliders.
Hence, for the benchmark scenario considered here and for an integrated luminosity of $\SI{100}{fb^{-1}}$, we expect $\mathcal{O}(1000)$ of those decays. Considering gluon fusion as the Higgs production channel, this number even increases to $\mathcal{O}(10^5)$.

\begin{table}[b]
	\centering
	\begin{tabular}{l|cccccccc}
	\toprule
	\multirow{2}{*}{Collider} & \multirow{2}{*}{\symhspace{0.2em}{$\delta_2\cdot 10^3$}} & \hspace{0.5em} & \multicolumn{2}{c}{$m_h=\SI{5}{GeV}$} & \hspace{0.5em} & \multicolumn{3}{c}{$m_h=\SI{10}{GeV}$} \\
	& & & \symhspace{0.5em}{$N_{4\tau}$} & \symhspace{0.5em}{$N_{6\tau}$} & & \symhspace{0.45em}{$N_{4\tau}$} & \symhspace{0.5em}{$N_{2b4\tau}$} & \symhspace{0.45em}{$N_{4b2\tau}$} \\
	\colrule
	\vspace{-0.9em}\\
	\multirow{2}{*}{LHC} & -2 & & 72 & 9270 & & 83 & \num{47100} & \num{61800} \\
	& 5 & &  \num{31500} & 8760 & & \num{41100} & \num{44400} & \num{58500} \\
	\colrule
	\vspace{-0.9em}\\
	\multirow{2}{*}{ILC} & -2 & & 1 & 84 & & 1 & 428 & 560 \\
	&  5 & & 288 & 80 & & 375 & 405 & 530 \\
	\botrule
	\end{tabular}
	\caption{Event numbers for the \SI{14}{TeV} LHC (\SI{250}{GeV} ILC) run with $\SI{300}{fb^{-1}}$ ($\SI{500}{fb^{-1}}$) of data. Higgs production via gluon fusion is assumed for the LHC. Model parameters were chosen as in Figure \ref{fig:pheno:cross-sections-5GeV}.}
	\label{tab:pheno:highMassNumbers}
\end{table}

As argued before, the higher light scalar's mass leads to a large variety of different final states being kinematically accessible.
In the context of \textit{two}-body scalar Higgs decays, a comprehensive survey of the conceivable final states and their detection prospects at the LHC was given in \cite{Curtin2014}.
Of course, all possible channels have their advantages and drawbacks in terms of overall rate and potential backgrounds.
The most promising ones are probably those containing pairs of taus and/or $b$-quarks.
Especially at the ILC, whose $\tau$ and $b$ tagging capabilities will be very good \cite{Baer2013}, those channels may be used for dedicated searches to observe or constrain three-body scalar Higgs decays in the high-$m_h$ regime.
But also at the LHC, a six-tau final state may well be within reach of the upcoming run.
Table \ref{tab:pheno:highMassNumbers} lists expected event numbers for some interesting channels.

%%%%%%%%%%%%%%%%%%%%%%%%%%%%%%%
%%%%%%%%  S E C T I O N :  Conclusion %%%%%%%%%
%%%%%%%%%%%%%%%%%%%%%%%%%%%%%%%

\section{Conclusion and outlook}
\label{sec:concl}
\noindent
%================================================
% introduction
In the present work, we studied processes in which the Standard Model (SM)-like Higgs boson found at the Large Hadron Collider (LHC) decays into multiple light scalars.
Whereas up to now, only decays into two scalars were considered relevant in the literature, we also included the three-body channel in our discussion and analyzed under which circumstances this extra process becomes important.

% model-independent scenarios with large \Gamma_3
We began by arguing that the scalar three-body decay channel opens as soon as there is a sufficiently light scalar particle that mixes with the physical LHC Higgs.
Employing a generic parametrization for the scalar potential after electroweak symmetry breaking and thus \textit{not depending on a particular model realization}, we then identified scenarios where significant three-body decay rates $\Gamma_3$ are obtained.
First, sizable cubic self-interactions of the light scalar lead to relatively large $\Gamma_3$.
However, these interactions are fundamentally limited by perturbative unitarity.
A numerical analysis showed that three-body scalar Higgs decays would therefore always be at least 1 order of magnitude less abundant than their two-body counterpart, if it was not for a further contribution to $\Gamma_3$.
This second contribution comes from renormalizable, tree-level contact interactions which \textendash\ if sufficiently strong \textendash\ can lead to three-body rates comparable to or even exceeding those of scalar two-body decays.
Importantly, a similar enhancement mechanism does not exist for scalar $n$-body Higgs decays with $n>3$.
Here, the contact interactions correspond to nonrenormalizable operators and are thus suppressed by some high mass scale.
At the same time, the limitations on cubic self\hyp{}interactions remain valid.

% model-specific analysis (conditions for large \Gamma_3)
In a next step, we considered the SM extended by a real singlet as a specific particle physics model with an enlarged scalar sector.
We demonstrated that there are regions in parameter space, where the SM-like Higgs decays with comparable rates into two and three singletlike scalars, respectively.
The three-body decay can even be more abundant than the two-body one, if the effective portal coupling which mediates the latter becomes anomalously small.
As a consistency check, we made sure that all parameter points under consideration comply with both current experimental and theoretical bounds.
Applying the model-independent discussion from above, we then identified strong quartic self-interactions in the singlet sector combined with non-negligible scalar mixing as the main source of large $\Gamma_3$.
But also in the absence of mixing, there can be three-body scalar Higgs decays,%
\footnote{Note, however, that this feature is special to the particular model under investigation and is, for instance, not true if the singlet sector exhibits a discrete $\mathds{Z}_2$ symmetry.}
whose rates are, however, always suppressed with respect to those of two-body decays.
The three-body channel is entirely negligible if the singlet sector is practically decoupled from the SM and exhibits extremely weak self-interactions.

% model-specific analysis (collider signatures)
Finally, we analyzed the prospects for measuring scalar three-body Higgs decays at the upcoming \SI{14}{TeV} LHC run and future electron-positron colliders like the ILC.
Distinguished by the particles that the light scalar can decay into, we separately discussed two different regimes for its mass.
On the one hand, we considered the low-mass region with singletlike scalars lighter than approximately $\SI{1}{GeV}$.
Here, the light scalar predominantly decays into pion and muon pairs.
In the low-mass regime, the smoking-gun signature for a direct observation of scalar three-body Higgs decays is therefore a signal with three pairs of collimated muons each having the same invariant mass, namely that of the light scalar.
This constitutes a very clean event topology with hardly any SM background.
We found that, depending on the model parameters, the six-muon process might be in reach of the upcoming LHC run with up to $\mathcal{O}(100)$ events to be expected.
However, it is unlikely to serve as a discovery channel of a hidden singlet sector, since the corresponding four-muon process is always more abundant.
Still, if a beyond-the-SM four-muon signal is observed, a dedicated search for six-muon events can be used to distinguish new physics scenarios.
In particular, scalar three-body Higgs decays may be the only way to measure or constrain self-interactions of a light scalar sector.
Due to their larger rates, also searches for six-particle final states containing both muon and pion pairs might be interesting for this purpose.
Considering searches at the ILC, the six-pion final state was found to be the most promising signature to observe the three-body channel in the low-mass region.

In the high-mass regime, on the other hand, light scalars were assumed to be heavier than the $B$-meson threshold at approximately \SI{5}{GeV}.
Depending on their actual mass, the light scalars will thus mainly decay into tau leptons, gluons and $b$- and $c$-quarks so that a large variety of six-particle final states is conceivable.
We found that three-body scalar Higgs decays should be accessible in both LHC and ILC, provided scalar mixing is not too small and singlet quartic self-interactions are sufficiently strong.
In contrast to the low-mass region, the three-body decays even may serve as a discovery channel of a light scalar sector, since there exist scenarios where the two-body channel is anomalously small and thus unobservable.

% Outlook
Although our results already seem very promising, a final answer to the question of whether scalar three-body Higgs decays may be observed in near-future collider runs must be given by dedicated Monte Carlo simulations.
In particular, the final-state particles' $p_\text{T}$-spectra are needed in order to reliably assess both trigger and detector acceptance for the individual channels.
Moreover, a thorough analysis of potential background processes will be crucial.
Still, already at this point, it would be interesting to perform a phenomenological study similar to ours also in the context of different beyond-the-SM theories with enlarged scalar sectors.
Interesting and already well-studied theories with firm theoretical justification include two\hyp{}Higgs\hyp{}doublet models with or without extra singlets, real and complex singlet-extensions with additional symmetry or models involving scalar triplets.
Most of these models do not only introduce new $CP$-even but also extra $CP$-odd mass eigenstates, which may be light as well.
Assuming that $CP$ is a good symmetry of the scalar sector at low energies, the detection of three-body Higgs decays could then additionally be used to rule out pseudoscalars as the observed new particles.

% Conclusion's conclusion
In conclusion, the present work demonstrates that scalar three-body Higgs decays are worth studying both from an experimental and a theoretical point of view.

%%%%%%%%%%%%%%%%%%%%%%%%%%%%%%%%
%%%%%%%%  A C K N O W L E D G E M E N T S  %%%%%%
%%%%%%%%%%%%%%%%%%%%%%%%%%%%%%%%

\section*{acknowledgements}
\noindent
%================================================
The authors would like to thank Pascal Humbert, Karl Jakobs, Tilman Plehn, and Kai Schmitz for valuable discussions and helpful comments on the manuscript.

%%%%%%%%%%%%%%%%%%%%%%%%%%%%%%%%
%%%%%%%%  A P P E N D I X  %%%%%%%%%%%%%%%
%%%%%%%%%%%%%%%%%%%%%%%%%%%%%%%%
\appendix

%%%%%%%%%%%%%%%%%%%%%%%%%%%%%%%%%%%%%
%%%%%%%%  S E C T I O N :  Phase space threshold functions %%%%
%%%%%%%%%%%%%%%%%%%%%%%%%%%%%%%%%%%%%

\section{Phase space threshold functions}
\label{app:phaseSpace}
\noindent
%================================================
For a compact formulation of the scalar multibody decay rates in Section \ref{sec:decays}, we introduced different threshold functions $\ell_i$, the exact forms of which are the subject of this appendix.
In contrast to the two-body case with threshold function
\begin{align*}
	\ell_2(x) = \sqrt{1-4x^2} \fineq{,}
\end{align*}
the phase space integration for three final-state particles can in general not be performed analytically.

The corresponding kinematic threshold functions $\ell_3$ for degenerate final-state masses have the following integral representation (see \eg \cite{Olive2014}):
\begin{align*}
	\ell_3^{(n,m)}(x) = 2 \int_{\epsilon^-}^{\epsilon^+} \!\! \dd \epsilon \int_{\eta^-}^{\eta^+} \!\! \dd \eta \left( \epsilon-x^2 \right)^{-n} \left( \eta-x^2 \right)^{-m} \fineq{}
\end{align*}
for $n,m \in \mathds{N}_0$. The integration boundaries are given by
\begin{align*}
	\epsilon^- & = 4x^2 \sepeq{,}
	\epsilon^+ = (1-x)^2 \fineq{,} \\
	\eta^\mp & = \frac{1}{4\epsilon} \biggl[ (1-x^2)^2 - \left( \lambda^{\nicefrac{1}{2}}(\epsilon,x^2,x^2) \pm \lambda^{\nicefrac{1}{2}}(1,\epsilon,x^2) \right)^2 \biggr] \fineq{,}
\end{align*}
where $\lambda(a,b,c) \equiv (a-b-c)^2-4bc$ is the K\"all\'en triangle function. For vanishing $m$ (or $n$) the $\eta$-integration can be performed and the threshold function simplifies to
\begin{align*}
	\ell^{(n)}_3(x) & = \int_{\epsilon^-}^{\epsilon^+} \!\!\! \dd \epsilon \; \frac{L_3(\epsilon, x^2)}{(\epsilon - x^2)^n} \fineq{}
\end{align*}
with integral kernel
\begin{align*}
	L_3(\epsilon, x^2) \equiv 2(\eta^+ - \eta^-)
	 = \frac{2}{\epsilon} \lambda^{\nicefrac{1}{2}}(\epsilon,x^2,x^2)
	 \lambda^{\nicefrac{1}{2}}(1,\epsilon,x^2) \fineq{.}
\end{align*}

The threshold functions are normalized such that $\ell_2(0) = \ell_3^{(0)}(0) = 1$. Note furthermore that all $\ell_i$ vanish at the respective production threshold, \ie $\ell_2(\tfrac{1}{2})=\ell_3^{(n,m)}(\tfrac{1}{3}) = 0$, hence their name.

%%%%%%%%%%%%%%%%%%%%%%%%%%%%%%%%%%%%%
%%%%%%%%  S E C T I O N :  Tree-level unitarity %%%%
%%%%%%%%%%%%%%%%%%%%%%%%%%%%%%%%%%%%%

\section{Tree-level unitarity}
\label{app:unitarity}
\noindent
%================================================
The present appendix is meant to complement the discussion on tree-level perturbative unitarity from Section \ref{sec:SMS:theoretical}. In particular, we provide details on the calculations whose results were used to set limits on some of the scalar couplings.

\begin{figure}[t]
	\centering
	\subfloat[]{\includegraphics[scale=0.36]{\figPath/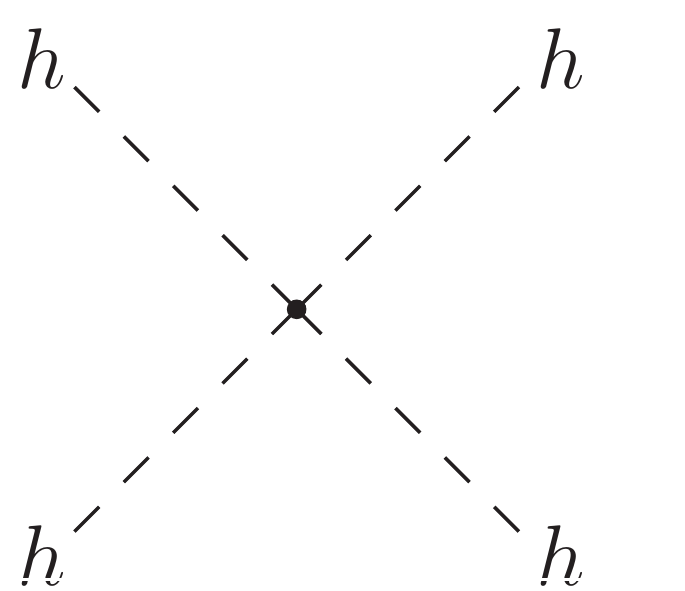}}%
	\hspace{4em}%
	\subfloat[]{\includegraphics[scale=0.36]{\figPath/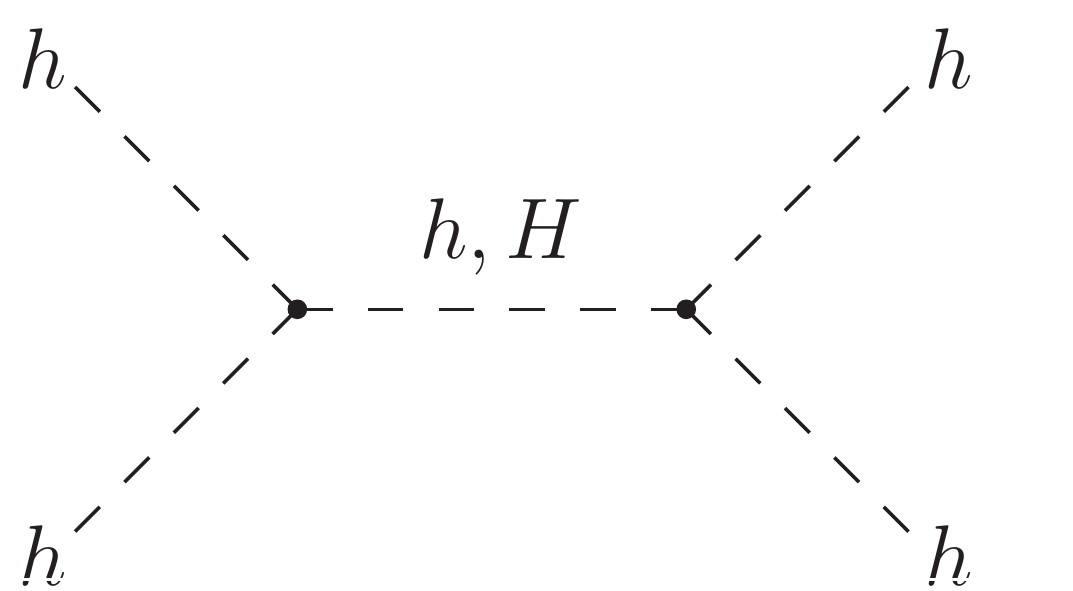}}%
	\subfloat{}%
	\caption{Feynman graphs contributing to light Higgs scattering $hh\to hh$ at tree-level. For (b) the corresponding $t$- and $u$-channel diagrams also exist but are not shown.}
	\label{fig:unitarity:hhhh}
\end{figure}

As mentioned in Section \ref{sec:SMS:theoretical}, the severest constraints on the model's parameter space originate from light Higgs elastic scattering, $hh\to hh$. The Feynman diagrams associated with this process at tree-level are displayed in Figure \ref{fig:unitarity:hhhh}. If $\mathcal{M}(s,\cos\vartheta)$ is the corresponding invariant matrix element, then the $j$th partial-wave amplitude is given by (for $j\geq0$)
\begin{align}
	a_j(s) = \frac{1}{32\pi} \int_{-1}^1 \!\!\dd \cos\vartheta \; P_j(\cos\vartheta) \mathcal{M}(s, \cos\vartheta) \fineq{.}
	\label{eq:unitarity:amplitudes}
\end{align}
In the following, we will only need the $s$-wave amplitude $a_0$ which can be obtained from equation \eqref{eq:unitarity:amplitudes} by using $P_0(\cos\vartheta)\equiv1$.
The relevant unitarity bound now reads
\begin{align}
	\absVal{\xi(s)\operatorname{Re}a_0(s)} \leq 1 \quad
	\sepeq{\text{with}} \xi(s) = \sqrt{1-\frac{4m_h^2}{s}} \fineq{,}
	\label{eq:unitarity:bound}
\end{align}
which must hold for all kinematically allowed values of the center-of-mass energy $\sqrt{s}\geq 2m_h$.
The various upper limits on $\kappa_3$, $\delta_2$ and $\kappa_4$ as indicated in equations \eqref{eq:SMS:unitarity_kappa3} to \eqref{eq:SMS:unitarity_kappa4} result from different energy ranges.

First, consider the situation in the vicinity of the Higgs pole, \ie $s\simeq m_H^2$. Here, the $s$-channel heavy Higgs-exchange diagram hits a resonance such that all other contributions will be negligible. The matrix element can therefore be approximated as
\begin{align*}
	\mathcal{M}(s\simeq m_H^2, \cos\vartheta) \simeq -\frac{4\kappa_{H2h}^2}{s-m_H^2 -\I m_H \Gamma_H} \fineq{.}
\end{align*}
The corresponding $s$-wave amplitude can be easily calculated, and one ends up with
\begin{align}
	\absVal{\operatorname{Re}\tilde{a}_0} = \frac{\kappa_{H2h}^2}{4\pi} \frac{\absVal{\delta s}}{\delta s^2 + m_H^2 \Gamma_H^2} \fineq{,}
	\label{eq:unitarity:a0-HiggsPole}
\end{align}
where $\delta s := s-m_H^2$ and we used $\xi\simeq 1$ for $m_h\ll m_H$. The above function exhibits a maximum at $\delta s_0 = m_H \Gamma_H$. Evaluating equation \eqref{eq:unitarity:a0-HiggsPole} at $\delta s_0$ and applying the unitarity bound \eqref{eq:unitarity:bound} gives
\begin{align}
	\frac{\kappa_{H2h}^2}{8\pi m_H\Gamma_H} \leq 1 \quad
	\sepeq{\Leftrightarrow} \absVal{\delta_2} \leq \sqrt{32\pi m_H \Gamma_H/v^2} \fineq{,}
	\label{eq:unitarity:bound-HiggsPole}
\end{align}
with the equivalence strictly holding in the $\theta\to 0$ limit.

Now, let us assume that we are far away from the Higgs pole at $s=m_H^2$. Then all diagrams with an internal $H$ propagator are suppressed by the large Higgs mass and thus will not contribute significantly. 
Additionally exploiting that $\Gamma_h/m_h \ll 1$, the matrix element can be written as
\begin{align*}
	\mathcal{M}&(s, \cos\vartheta) \simeq -24\lambda_{4h} - 36\kappa_{3h}^2 \Biggl[ \frac{1}{s-m_h^2} \\
	& - \frac{1}{(1-\cos\vartheta)s\xi^2/2 + m_h^2} - \frac{1}{(1+\cos\vartheta)s\xi^2/2+m_h^2} \Biggr] \fineq{.}
\end{align*}
The integration over the polar angle $\vartheta$ in \eqref{eq:unitarity:amplitudes} can be performed analytically ultimately resulting in
\begin{align}
	\absVal{\operatorname{Re}\tilde{a}_0(y)} = \frac{3\xi(y)}{2\pi} \Biggl| \lambda_{4h}
	- \frac{3\kappa_{3h}^2}{2m_h^2} \cdot g(y) \Biggr| \fineq{,}
	\label{eq:unitarity:a0-notHiggsPole}
\end{align}
where we defined $y:=2m_h/\sqrt{s}$ such that $\xi^2(y)={1-y^2}$. In the kinematically allowed range, $y$ runs from zero at asymptotically large energies to one at threshold.
The function $g(y)$ in equation \eqref{eq:unitarity:a0-notHiggsPole} is given by
\begin{align}
	g(y) := - \frac{y^2}{4-y^2} + \frac{y^2}{2\xi^2(y)} \log\left( 1+\frac{4\xi^2(y)}{y^2} \right) \fineq{.}
\end{align}
Here, the first term stems from $s$-channel $h$-exchange, whereas the second term includes contributions from both $t$- and $u$-channel diagrams.
It is straightforward to show that $g(y)$ is non-negative and strictly monotonously increasing for all $y\in[0,1]$. In other words, $t$- and $u$-channel amplitudes dominate.
Moreover, $g(y)$ is zero only in the high-energy limit $y\to0$, where both terms vanish individually and also $\xi\to 1$. Consequently, applying equation \eqref{eq:unitarity:bound} for asymptotically large energies gives
\begin{align*}
	\frac{3 \lambda_{4h}}{2\pi} \leq 1 \quad
	\sepeq{\Leftrightarrow} \kappa_4 \leq \frac{8\pi}{3} \fineq{,}
\end{align*}
where the equivalence holds for small scalar mixing angles and $\lambda_{4h}$ as well as $\kappa_4$ must be non-negative due to vacuum stability reasons.

Furthermore, the function in equation \eqref{eq:unitarity:a0-notHiggsPole} exhibits a local maximum at some value $y_0 \in (0,1)$ provided the ratio $\tilde{\kappa}_{3h}:=\kappa_{3h}/m_h$ is sufficiently large. In a first step, analyzing equation \eqref{eq:unitarity:a0-notHiggsPole} for negligible $\lambda_{4h}$, one finds a maximum at $y_0=\num{0.85}$. Requiring that the associated function value satisfies equation \eqref{eq:unitarity:bound} gives the constraint
\begin{align}
	\num{0.37}\cdot \tilde{\kappa}_{3h}^2 \leq 1 \quad
	\sepeq{\Leftrightarrow} \kappa_3\leq \num{4.9}\cdot m_h \fineq{,}
	\label{eq:unitarity:bound-Threshold}
\end{align}
where we used the small-$\theta$ expansion from equation \eqref{eq:SMS:smallThetaExp}.
Sizable values for $\lambda_{4h}$ turn out to relax the above bound. Numerical evaluation shows that, for instance, $\lambda_{4h}=\tfrac{1}{2}$ leads to a modified upper limit of $\tilde{\kappa}_{3h}\lesssim \num{5.2}$.

%%%%%%%%%%%%%%%%%%%%%%%%%%%%%%%%
%%%%%%%%  B I B L I O G R A P H Y  %%%%%%%%%%%%
%%%%%%%%%%%%%%%%%%%%%%%%%%%%%%%%
\bibliographystyle{../../references/bibstyle}
\bibliography{../../references/References-MPIK}

\end{document}